# A new method to identify elastic wave propagation mode and polarized bandgaps in periodic solid media

Maria Jose Carrillo Munoz and Bhisham Sharma[*]

Department of Aerospace Engineering, Wichita State University, Wichita, KS 67260-042, USA

* Corresponding author: bhisham.sharma@wichita.edu

**Abstract**

We present a new computational method for the accurate identification of the propagation modes and polarizations of elastic waves propagating in periodic solid structures and metamaterials. The method uses the eigenvectors calculated at each propagating wave solution eigenfrequency to identify the contribution of each translational and rotational component of the total mass in motion. We use this information to identify the dominant wave propagation mode by defining a relative effective modal mass vector. Further, we associate each wave solution with its correct polarization by defining a new polarization factor that quantifies the relative orientation between the wave propagation and lattice motion directions and provides a positive numerical value between 0 (pure *S*-wave) and 1 (pure *P*-wave). Further, we suggest a graphical representational scheme for an easier visualization of the wave polarization within traditional dispersion plots. The method is validated by comparing our predictions against previously published results for various elastodynamic problems. Finally, we use the proposed method to analyze the effect of various lattice and structural parameter perturbations on the elastic wave propagation and bandgap behavior of a square planar beam lattice and show the emergence of previously unobserved dynamic characteristics, including various polarized bandgaps, fluid-like behavior, and ultralow-frequency *SH*- and *SV*-bandgaps that extend to 0 Hz. The proposed method provides an alternative computational approach to the typically employed visual mode inspection technique and provides a robust technique for analyzing the elastic wave response of periodic solid media.

## 1. Introduction

Interest in phononic media and metamaterials stems from their novel bulk properties and from the possibility of controlling elastic energy flow through structures [1-3]. Classically, research on these material systems has been focused on the emergence of wave attenuation bandgaps—frequency zones within which incident waves are spatially attenuated—as a function of their underlying substructure. In phononic media, this attenuation occurs because of substructural periodicity and the resultant Bragg scattering effects [4]; in metamaterials, the attenuation occurs due to localized substructural resonances sequestering the incident elastic energy [3, 5]. While metamaterials do not strictly require a periodic substructure, this assumption is frequently made for the convenience of analysis and eventual fabrication, leading to the appearance of both Bragg and local resonance attenuation effects [6-8].

The analysis of elastic wave propagation through such structures is mathematically complicated, vis-à-vis their photonic counterparts [5], because of the multiple wave modes or polarizations supported by elastic media [9, 10], their propensity to couple at higher frequencies [11-13], and their interconversion during reflections at the substructural interfaces [14, 15]. Polarization is defined as the relative orientation between the wave propagation direction—conveniently described by the wave vector orientation—and the oscillation direction of the elastic particles [16, 17]. An unbounded homogenous elastic medium supports the propagation of three uncoupled wave polarizations [10]: one *P*-wave—also called dilatational, irrotational, longitudinal, compressional, voluminal, or primary (*P*) wave—wherein the particles oscillate parallel to the wave propagation direction; and two *S*-waves—also called distortional, equivoluminal, rotational, transversal, shear, or secondary (*S*) waves—wherein the particles oscillate perpendicular to the wave propagation direction. The two *S*-waves are further distinguished as *SH*- or *SV*-waves if the particle motion occurs in-plane (horizontal) or out-of-plane (vertical), respectively, relative to the plane defined by the particle oscillation, wave propagation direction, and the global coordinate choice. Heterogenous or anisotropic unbounded media support additional coupled waveforms with polarizations neither parallel nor perpendicular to the propagation direction [18, 19]. Depending on the dominant polarization, such coupled waves are classified as quasi-*P* or quasi-*S* waves; lacking a dominant polarization, such waves may be classified as hybrid [20]. Waveforms in bounded media are more commonly classified as longitudinal (equivalent to *P*-waves), flexural or bending or transverse (equivalent to *SV*-waves), shear (equivalent to *SH*-waves), and torsional waves where the particle motion comprises twisting or rotation about the propagation direction [10].

Depending on their underlying substructure—periodicity, property contrast ratios, symmetry, and connectedness all play an important role—phononic media and metamaterials exhibit distinct bandgaps that can be classified either based on their generation mechanism as Bragg [4] or local resonance bandgaps [3], as explained earlier, or based on the waveform types they attenuate. Polarized bandgaps [21, 22] restrict the propagation of waves only with specific polarizations; for example, a *P*-wave bandgap only attenuates incident *P*-waves while waves with other polarizations propagate within that frequency range without spatial attenuation. The superposition of all polarized bandgaps within the same frequency range results in a complete or full bandgap within which all waves, irrespective of their polarization, are attenuated [21]. In multidimensional structures, the bandgaps are further classified depending on their directional nature. Directional bandgaps [23-25] are frequency ranges where waves only along specific wave vector directions are attenuated; waves with other incident angles or propagation directions propagate unattenuated through the structure. The overlap of such directional bandgaps in all $4\pi$ radians results in absolute or omnidirectional bandgaps [26]. Thus, depending on their polarization and directionality, one may classify bandgaps as polarized-directional, polarized-omnidirectional, complete-directional, or complete-omnidirectional. Wave filtering, because of directional or polarized-directional bandgaps, forces incident waves to propagate in the other permitted directions. This phenomenon has been harnessed by researchers to show the possibility of beaming or steering elastic waves in specific directions [24, 25] and is also of interest for the inverse problem of non-destructively identifying damage by monitoring elastic energy flow [27, 28]. Recently, researchers have also observed the appearance of polarization anomalies because of substructural heterogeneities [29-31], where the typically slower *S*-waves propagate faster than the *P*-waves.

These phenomena are traditionally studied by deriving the spectral or dispersion relations of the structure under consideration. Over the years, numerous analytical and numerical techniques have been used to obtain these relationships, including using equivalent spring-mass analogs [4, 32], Hamiltonian energy method [33, 34], space-harmonic method [35], receptance technique [36, 37], transfer matrix method [6, 38], phased array method [7, 39], finite element method [25, 40-42], and the spectral element method [43, 44]. In general, these methods use a representative unit cell in conjunction with the Floquet-Bloch periodicity boundary condition to extract the wave number-frequency relationships of waves propagating through the global structure [45]. While analytical methods become intractable as the substructural complexity increases, the finite element method provides a reliable route for analyzing such structures. Regardless of the method choice, the obtained spectral curves or surfaces provide information about the dispersion behavior of the propagating waves. Omnidirectional and complete bandgaps are then easily identified as frequency regions with no real-valued solutions. However, associating each dispersion curve or surface with its associated polarization and consequently identifying individual polarized and polarized-directional bandgaps requires further analysis. These polarized bandgaps are most frequently identified by observing the mode shapes at the bandgap edge frequencies at the high-symmetry irreducible Brillouin zone (IBZ) points [1, 21, 22, 46]. While this is easily done at lower frequencies where the wave polarizations are usually uncoupled, manually identifying the polarizations gets complicated for coupled waves. Further, curve veering—described by Manconi and Mace [12] as a phenomenon where *"two or more eigenvalue loci of a system with a varying parameter veer away and diverge instead of crossing"*—and the possible continuous variation of polarization as the wave vector sweeps the IBZ [17], further complicate the accurate wave polarization identification using mode shape observation at high-symmetry points. Some researchers have employed alternative methods to identify polarizations. Focusing on sagittal acoustic waves, Manzanares-Martinez and Ramos-Mendieta [17] adopted a strain energy balance approach and obtained the averaged longitudinal and transversal displacement contributions for one-dimensional phononic crystal unit cells; for two-dimensional cases, the authors avoid defining a local wave vector by instead calculating the average compression and shear contributions and approximating them to longitudinal and transversal vibrations. Similarly, Achaoui et al. [47] used the integrals of displacement field components as representative of wave polarization without considering the wave vector direction. A different approach, accounting for the wave vector direction, was adopted recently by Bacigalupo and Lepidi [20], who proposed a family of non-dimensional polarization factors to quantify the wave polarization and energy flow in periodic beam lattice materials and showed the coincidence between the energy and group velocities. To predict out-of-plane wave beaming in two-dimensional lattices, Zelhofer and Kochmann [25] distinguished the in-plane and out-of-plane wave modes using a scalar in-plane ratio calculated using the mass normalized eigenvectors. Lee et al. [30] identified the polarization characteristics of a double-slit metamaterial showing polarization anomalies by calculating the absolute value of the relative angle between the polarization orientation and the wave vector orientation.

In this work, we propose a systematic approach to identify wave propagation modes and to accurately associate the dispersion curves with their respective polarizations. We adapt the concept of the modal participation factor [48], commonly used in vibration modal analysis [49, 50], to quantify the relative participation of each directional motion component to the wave mode associated with each dispersion solution. The relative orientation between the wave propagation and particle motion direction is then calculated by defining a polarization factor that provides a positive numerical value between 0 and 1, where 0 indicates a pure *S*-wave, 1 indicates a pure *P*-wave, and the interim values indicate quasi- or hybrid polarizations. Further, we suggest a graphical representational scheme for easier visualization of the wave polarization within traditional dispersion plots. Though our focus here is on utilizing existing eigenvector data that is calculated while numerically extracting the dispersion curves, the method is easily adaptable for use with other techniques that provide eigenvector data. The novelty of the proposed method lies in its ability to clearly elucidate the wave propagation mode and polarizations underlying each dispersion solution; this clarity enables the identification of previously unobserved emergent behaviors in widely studied lattice geometries. The following section provides a background on the current computational method used to obtain dispersion curves and motivates the presented work by discussing the inherent issues within the commonly used method of visual mode shape analysis at the high-symmetry points. Then, the proposed approach and data visualization schema are presented and validated by comparing our results against previously published results for various elastodynamic problems. Finally, we use the proposed approach to study the elastic wave propagation behavior of a square planar beam lattice structure and show the existence of multiple polarized and directional bandgaps and analyze the effect of various lattice and structural parameter perturbations on the individual bandgaps.

## 2. Background and Motivation

### *2.1. Elastic wave dispersion relations*

Elastic wave propagation in phononic structures and metamaterials with substructural periodicity is conveniently studied by applying the Floquet-Bloch boundary conditions on a representative unit cell [45]. The time-harmonic displacement field within an infinite periodic structure can be represented as:

$$\mathbf{u}_A(\mathbf{r}, t) = \mathbf{U}_A(\mathbf{r}) \, e^{-i\omega t}, \tag{1}$$

where $\mathbf{U}_A$ is the complex displacement amplitude at point *A* with a position vector $\mathbf{r}$, *t* is the time variable, $\omega$ is the wave frequency, and $i = \sqrt{-1}$. According to the Floquet-Bloch theory, the response of any point *A* in a periodic structure is related to the response of the corresponding spatially periodic point *A′* with position vector $\mathbf{r}'$ through the complex wave vector $\mathbf{k}$ as [4, 45]:

$$\mathbf{U}_{A'}(\mathbf{r}') = \mathbf{U}_A(\mathbf{r}) \, e^{-i\mathbf{k}(\mathbf{r}'-\mathbf{r})} \tag{2}$$

In this representation, the real component of the complex wave number is the phase difference between points $A$ and $A'$, while the imaginary component—also called the wave attenuation factor—is the wave amplitude decay rate between the two points; the wave vector direction indicates the propagation direction of the elastic wave. For non-dissipative linear materials, $\mathbf{k}$ is purely real for the propagating waves and becomes purely imaginary within wave attenuation bandgaps—frequency regions where waves are spatially attenuated and do not propagate through the structure. The dispersion behavior of the propagating waves and the existence of bandgaps is studied by applying the Floquet-Bloch boundary condition on the unit cell and then calculating the eigenfrequency solutions while sweeping over the wave vectors of interest. This approach, frequently referred to as the $\omega(\mathbf{k})$ approach [1, 46], leverages the theorem that a structure which is periodic in the physical space with a periodicity $L$ is also periodic in the wave vector space—also called the reciprocal space—with periodicity $2\pi/L$. Thus, a unit cell with periodicity $L_1$, $L_2$, and $L_3$ along the *1*-, *2*-, and *3*-axis results in a reciprocal space with wave number periodicity $k_1 \to 2\pi/L_1$, $k_2 \to 2\pi/L_2$, and $k_3 \to 2\pi/L_3$. The unit cell representing this resultant reciprocal space is called the Brillouin zone [45]. By choosing the reciprocal unit cell to be centered around the point $\mathbf{k} = 0$ and allowing only positive $\mathbf{k}$ values, we can reduce the computational wave vector domain to $k_1 \to [0, \pi/L_1]$, $k_2 \to [0, \pi/L_2]$, and $k_3 \to [0, \pi/L_3]$. This reduced domain—called the First Brillouin Zone—can be further shrunk by exploiting its symmetry to get the Irreducible Brillouin Zone (IBZ) [45]. As the name suggests, the IBZ is the smallest possible reciprocal unit cell which provides complete information about a periodic structure's spectral behavior. Thus, the $\omega(\mathbf{k})$ approach involves sweeping over the IBZ while solving for the real-only frequencies allowed by the structure.

For complicated geometries, the above technique is frequently implemented using the finite element (FE) method, where the unit cell is modeled using an appropriate mesh and the boundary nodes are constrained using Eq. (2). While some commercial FE software (e.g., Comsol Multiphysics) allow the direct application of complex-valued boundary conditions, others (e.g., Abaqus) only allow the application of real-valued displacements. Åberg and Gudmundson [42] showed that this limitation can be overcome by separating the model into two identical 'real' and 'imaginary' mesh parts and by formulating the total complex-valued displacement at a lattice point $A$ as:

$$\mathbf{U}_A = \mathbf{U}_{A_{\text{re}}} + i\, \mathbf{U}_{A_{\text{im}}} \tag{3}$$

where $\mathbf{U}_{A_{\text{re}}}$ and $\mathbf{U}_{A_{\text{im}}}$ are the displacements of the corresponding nodes at point $A$ of the real and imaginary models, respectively. The complex Floquet-Bloch boundary conditions in Eq. (2) are then implemented by applying Euler's equation to get:

$$\mathbf{U}_{A'_{\text{re}}} = \mathbf{U}_{A_{\text{re}}} \cos(\mathbf{k}(\mathbf{r}' - \mathbf{r})) + \mathbf{U}_{A_{\text{im}}} \sin(\mathbf{k}(\mathbf{r}' - \mathbf{r})) \tag{4}$$

$$\mathbf{U}_{A'_{\text{im}}} = \mathbf{U}_{A_{\text{re}}} (-\sin(\mathbf{k}(\mathbf{r}' - \mathbf{r}))) + \mathbf{U}_{A_{\text{im}}} \cos(\mathbf{k}(\mathbf{r}' - \mathbf{r})) \tag{5}$$

The eigenfrequencies corresponding to the wave vector values within the IBZ are then extracted using a linear perturbation analysis procedure. The equations of motion for the free vibration of the unit cell with boundary nodes constrained using Eq. (4) and Eq. (5) are:

$$\mathbf{M}\ddot{\mathbf{u}} + \mathbf{K}\mathbf{u} = 0 \tag{6}$$

where **u** and **ü** are the displacement and acceleration vectors, with **ü** signifying two time derivatives of **u**, and **M** and **K** are the global mass and stiffness matrices, respectively. Assuming the solution to be the time-harmonic displacement field given in Eq. (1) leads to the eigenvalue problem:

$$(\mathbf{K} - \omega^2 \mathbf{M})\mathbf{U} = \mathbf{0} \tag{7}$$

By introducing a constraint matrix **Q**, the equations can be rewritten in terms of the displacements of the unconstrained internal nodes, $\mathbf{U}_{int}$, and the master boundary nodes, $\mathbf{U}_{bdry}$, as:

$$\begin{bmatrix} \mathbf{U}_{int_{re}} \\ \mathbf{U}_{bdry_{re}} \cos(\mathbf{k}(\mathbf{r'} - \mathbf{r})) + \mathbf{U}_{bdry_{im}} \sin(\mathbf{k}(\mathbf{r'} - \mathbf{r})) \\ \mathbf{U}_{bdry_{re}} \\ \mathbf{U}_{int_{im}} \\ \mathbf{U}_{bdry_{re}} (-\sin(\mathbf{k}(\mathbf{r'} - \mathbf{r}))) + \mathbf{U}_{bdry_{im}} \cos(\mathbf{k}(\mathbf{r'} - \mathbf{r})) \\ \mathbf{U}_{bdry_{im}} \end{bmatrix} = \mathbf{Q} \begin{bmatrix} \mathbf{U}_{int_{re}} \\ \mathbf{U}_{bdry_{re}} \\ \mathbf{U}_{int_{im}} \\ \mathbf{U}_{bdry_{im}} \end{bmatrix} \tag{8}$$

where the matrix **Q** is defined as:

$$\mathbf{Q} = \begin{bmatrix} \mathbf{I} & 0 & 0 & 0 \\ 0 & \cos(\mathbf{k}(\mathbf{r'} - \mathbf{r})) & 0 & \sin(\mathbf{k}(\mathbf{r'} - \mathbf{r})) \\ 0 & \mathbf{I} & 0 & 0 \\ 0 & 0 & \mathbf{I} & 0 \\ 0 & -\sin(\mathbf{k}(\mathbf{r'} - \mathbf{r})) & 0 & \cos(\mathbf{k}(\mathbf{r'} - \mathbf{r})) \\ 0 & 0 & 0 & \mathbf{I} \end{bmatrix} \tag{9}$$

The resulting eigenvalue problem is a function of the wave vector and is given as:

$$\left( \begin{bmatrix} \mathbf{K} & 0 \\ 0 & \mathbf{K} \end{bmatrix} - \omega^2 \begin{bmatrix} \mathbf{M} & 0 \\ 0 & \mathbf{M} \end{bmatrix} \right) \mathbf{Q} \begin{bmatrix} \mathbf{U}_{int_{RE}} \\ \mathbf{U}_{bdry_{RE}} \\ \mathbf{U}_{int_{IM}} \\ \mathbf{U}_{bdry_{IM}} \end{bmatrix} = \begin{bmatrix} 0 \\ 0 \\ 0 \\ 0 \end{bmatrix} \tag{10}$$

Solving this equation provides the eigenfrequency solutions corresponding to the specific wave vectors. Graphical representation of these solutions as $\omega$ v/s **k** plots provides a clearer understanding of the structure's dispersion behavior and the presence of bandgaps. For a general three-dimensional case, the solutions can be plotted for the entire IBZ volume to obtain the dispersion volumes; plotting over a two-dimensional grid covering the two-dimensional IBZ provides dispersion surfaces; plotting along a one-dimensional path—typically chosen to traverse the IBZ boundaries—provides dispersion curves. The obtained information can be further processed to calculate the group and phase velocities and to study the structure's wave directionality behavior [24, 25, 51].

### *2.2. Elastic wave polarization*

As an illustrative example, consider the dispersion plot for an infinite, planar square beam lattice, shown in Fig. 1, with six degrees-of-freedom, obtained using the procedure described in section 2.1. We model the lattice in Abaqus CAE as an interconnected network of three-dimensional deformable wires, meshed using the shear-flexible beam elements with quadratic interpolation (B32). We assume that the unit cell lies in the *1-2* plane and that its lattice constants along the *1-* and *2*-axis are $L_1 = 1$ m and $L_2 = 1$ m, respectively. The lattice elements are assumed to have a square cross-section of 77 mm thickness. The dispersion curves obtained along the IBZ path O-X-M-O are shown in Fig. 1, where we normalize the frequency axis with respect to the frequency of the first in-plane bending mode occurring at the high-symmetry point X ($k_1 = \pi/L_1, k_2 = 0$). The mode shapes at specific dispersion point locations are also shown.

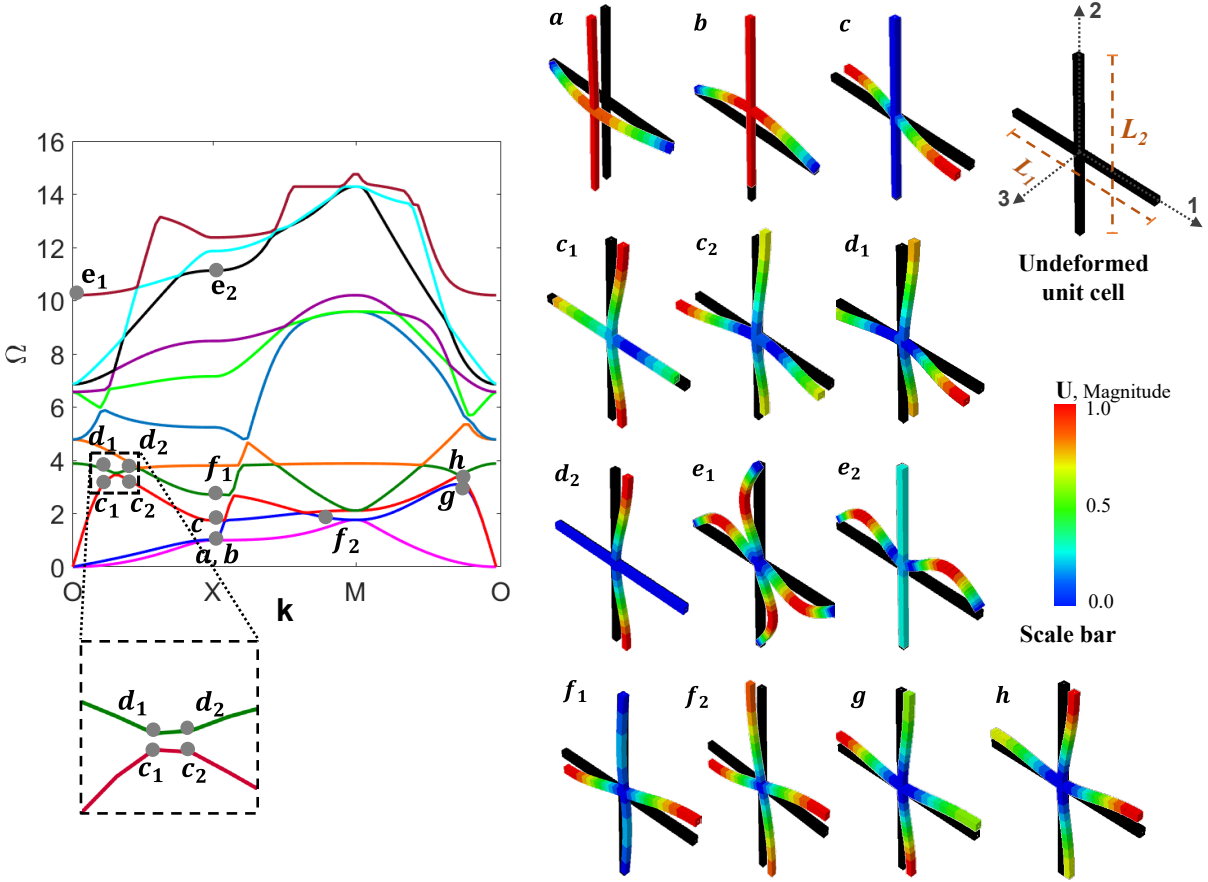

**Fig. 1.** Dispersion curves for a square beam lattice with six degrees-of-freedom and the unit cell mode shapes at the specified dispersion points. The inset shows the curve veering phenomena occurring between the 3$^{rd}$ and 4$^{th}$ eigenfrequency curves.

Each continuous eigenfrequency curve in the dispersion plots, shown here using distinct colors, represents a solution to Eq. (6), and each eigenfrequency represents a specific permitted wave propagation mode or polarization at that frequency. However, each continuous curve does *not* correspond to a single propagation mode; one must also pay attention to the associated eigenvectors to determine this information. Frequently, the associated polarizations are identified by visually inspecting the mode shapes at each eigenfrequency; to identify the attenuation bandgaps, the attention is usually restricted to the bandgap start and stop eigenfrequencies [21, 52]. For example, the mode shapes at the points *a*, *b*, and *c* along the high-symmetry point X show that their associated eigenfrequency curves, shown here in pink, blue, and red, respectively, are the propagating wave solutions for the first *SV*-, first *SH*-, and second *SV*-waves, respectively. However, further inspection of the mode shapes at points $c_1$ and $c_2$ along the third eigenfrequency (red) curve shows that while point $c_2$ is indeed the second *SV*-wave, point $c_1$ is a *P*-wave mode. This polarization change occurs because of the curve or mode veering phenomena [12, 13, 53] shown in the inset, where the dispersion curves associated with the *P*- and *SV*-waves abruptly switch polarizations and veer away and diverge instead of crossing over each other. Here, the eigenfrequency curves containing points $c_1$ and $c_2$, and the points $d_1$ and $d_2$ switch polarizations, as seen in the similarity of the mode shapes at points $c_1$ and $d_2$, and points $c_2$ and $d_1$. Such curve veering effects can lead to the incorrect identification of the wave polarizations associated with each dispersion curve. Another source of error in the visual mode shape inspection approach stems from the mode conversion effects, where the polarization associated with a single dispersion curve might vary continuously as a function of the wave vector [17]. For the square lattice, such mode conversion effects are seen in the $11^{th}$ dispersion curve along O-X: the curve starts ($e_1$) as a hybrid wave mode with similar contributions by the in-plane flexural motion in the vertical (*P*-mode) and horizontal (*SH*-mode) lattice elements but slowly coverts to an *SH*-wave mode, where the horizontal element's in-plane flexural motion is the dominant response, as seen at point $e_2$. A similar mode conversion occurs in the $3^{rd}$ dispersion curve along the X-M path: the curve begins at point $f_1$ with horizontal element dominated flexural motion along the *2*-direction and gradually converts to a hybrid mode with in-plane flexural motion of the horizontal and vertical elements along the *2*- and *1*-directions, respectively. Besides the mode veering and conversion effects, the coupled mode shapes underlying the waves propagating in directions unaligned with lattice element orientations, as seen at point *g*, further complicate the accurate identification of the wave polarizations. Consequently, while the complete and omnidirectional bandgaps are easily identified as frequency regions with no real eigen solutions, relying on visual mode shape inspection can easily result in the misidentification of polarized bandgaps.

Here, we propose an alternative method for associating the dispersion curves with their respective polarizations. For a FE mesh, every eigenfrequency is associated with an eigenvector at each node. The proposed method uses these eigenvectors to calculate the relative effective translational and rotational masses in motion along the different coordinate directions at each eigenfrequency. This information is then used to identify the structural vibration mode associated with the wave propagation at each frequency. While this is sufficient for the correct polarization identification of waves with wave vectors oriented along the structural elements, we propose the use of a polarization factor to ensure the accurate polarization identification of waves with different wave vector orientations. The following section explains the proposed method.

## 3. Method

### *3.1. Propagation mode identification*

The method described in section 2.1 provides the eigenfrequencies corresponding to the specific wave vectors of interest. Each eigenfrequency in the $n^{th}$ dispersion curve, $\omega^n$, is associated with an eigenvector, $\mathbf{U}^n$, at each FE node. The six eigenvector components $\mathbf{U}^n_j|_{j=1,2,..6}$ —where $\mathbf{U}^n_j|_{j=1,2,3}$ are associated with the three translational components $u_1^n, u_2^n, u_3^n$ along the *1, 2,* and *3* coordinate directions, respectively, and $\mathbf{U}^n_j|_{j=4,5,6}$ are associated with the three rotational components $\theta_1^n, \theta_2^n, \theta_3^n$ about the *1, 2,* and *3* coordinate directions, respectively—provide information about the relative motion of each FE node at $\omega^n$. The relative influence of each eigenvector component on the overall mode shape can be estimated by calculating the associated modal participation factor, $\mathbf{\Gamma}^n$, defined as:

$$\mathbf{\Gamma}^n = \frac{\mathbf{U}^{nT}\mathbf{M}\mathbf{R}}{\mathbf{m}^n} \tag{11}$$

where $\mathbf{M}$ is the global mass matrix, and $\mathbf{R}$ and $\mathbf{m}^n$ are the influence matrix and generalized mass, defined respectively as:

$$\mathbf{R} = \begin{pmatrix} 1 & 0 & 0 & 0 & (z-z_0) & -(y-y_0) \\ 0 & 1 & 0 & -(z-z_0) & 0 & (x-x_0) \\ 0 & 0 & 1 & (y-y_0) & -(x-x_0) & 0 \\ 0 & 0 & 0 & 1 & 0 & 0 \\ 0 & 0 & 0 & 0 & 1 & 0 \\ 0 & 0 & 0 & 0 & 0 & 1 \end{pmatrix} \tag{12}$$

$$\mathbf{m}^n = \mathbf{U}^{nT}\mathbf{M}\mathbf{U}^n \tag{13}$$

Here, $\mathbf{R}$, represents the displacements resulting from any static unit ground displacement or rotation, where *x, y, z* are the nodal coordinate components, and $x_0, y_0, z_0$ are the coordinates of the center of rotation along the *1-, 2-,* and *3*-axis, respectively. Note that by defining it this way, $\mathbf{R}$ degenerates into the identity matrix when the center of rotation coincides with the center of gravity.

While the modal participation factor is routinely used during modal analysis to understand the participation of individual normal modes to the overall dynamic state of a structure at a given frequency [54-58], here, we propose using it to understand the relative participation of each directional motion component to the mode shape associated with a given eigenfrequency. Thus, the modal participation factor component $\Gamma_j^n$ describes the contribution of the *j*[th] directional motion component to the overall mode shape at $\omega^n$. The modal participation factor can then be used to quantify the portion of the translational (*j* = 1, 2, 3) or rotational (*j* = 4, 5, 6) mass along each direction by calculating the modal effective mass as:

$$\mathbf{m}^n_{eff} = \mathbf{\Gamma}^{n2}\mathbf{m}^n \tag{14}$$

For convenience of comparison, we normalize the translational and rotational effective mass components as:

$$c_j^n = \frac{m^n_{eff\,j}}{M}, \quad \text{for } j = 1, 2, \text{ and } 3 \text{ (translational mass)} \tag{15}$$

$$c_j^n = \frac{m^n_{eff\,j}}{I_j}, \quad \text{for } j = 4, 5, \text{ and } 6 \text{ (rotational mass)} \tag{16}$$

where *M* and $I_j$ are the total translational and rotational mass of the structure. Note that while the translational mass is direction independent, the rotational mass (or moment of inertia) is directionally dependent and must be calculated about the appropriate axis. To better visualize the *relative* contribution of each effective mass component to the overall modal mass at $\omega^n$, we further normalize them with respect to the total translational and rotational effective masses as:

$$q_{t_j}^n \big|_{j=1,2,3} = \frac{c_j^n}{\sum_{j=1}^{3} c_j^n} \quad (17)$$

$$q_{r_j}^n \big|_{j=4,5,6} = \frac{c_j^n}{\sum_{j=3}^{6} c_j^n} \quad (18)$$

where $q_{t_j}^n$ and $q_{r_j}^n$ are the relative effective translational and rotational mass components along the $j^{th}$ direction at $\omega^n$, respectively. This additional normalization provides relative effective mass component values bounded between 0 and 1, where 0 indicates no mass motion along that coordinate direction and 1 indicates that all the mass is moving in that direction.

For clarity, consider locations $c_1$, $c_2$, $d_1$, and $d_2$ in the 3$^{rd}$ and 4$^{th}$ dispersion curves for the square planar lattice shown in Fig. 1 as an example. The wave vector direction for all four points is along the *1*-axis. The modal effective mass vectors and the relative effective modal mass vectors for these four points calculated using the above procedure are given in Table 1, where $\omega^n$ is the frequency and $\Omega^n$ is the normalized frequency of each location. At $c_1$ and $d_2$, the dominant motion of the structural mass, as predicted by the modal mass vectors, is primarily the in-plane displacement along the 1-axis; conversely, the motion at $c_2$ and $d_1$ is dominated by displacement along *3*-axis—i.e., out-of-plane—and rotation about the *2*-axis, indicating that both these points are associated the out-of-plane propagation mode. These predictions, verified by observing the mode shapes provided in Fig. 1, help identify the correct propagation mode at these four points affected by the curve veering phenomena. Similarly, the motion at locations *g* and *h*, studied further in Section 3.2, are also identified as in-plane. The modal effective mass vectors and the relative effective modal mass vectors for all locations shown in Fig. 1 are provided in the Appendix, Table A1.

**Table 1.** The calculated modal effective mass vectors, relative effective modal mass vectors, and the identified propagation mode for the specific points shown in Fig. 1.

| Location | Normalized frequency $\Omega^n$ | Modal effective mass vectors $\mathbf{m}_{eff}^n$ | Relative effective translational mass $\mathbf{q}_t^n = \langle q_{t_1}^n, q_{t_2}^n, q_{t_3}^n \rangle$ | Relative effective rotational mass $\mathbf{q}_r^n = \langle q_{r_4}^n, q_{r_5}^n, q_{r_6}^n \rangle$ | Propagation mode |
|---|---|---|---|---|---|
| $c_1$ | 3.41 | $\langle 14.4, 0, 0, 0, 0, 0 \rangle$ | $\langle 1, 0, 0 \rangle$ | $\langle 0.035, 0.996, 0.004 \rangle$ | In-plane |
| $c_2$ | 3.40 | $\langle 0, 0, 0.183, 0, 0.292, 0 \rangle$ | $\langle 0, 0, 1 \rangle$ | $\langle 0, 1, 0 \rangle$ | Out-of-plane |
| $d_1$ | 3.47 | $\langle 0, 0, 0.125, 0, 0.246, 0 \rangle$ | $\langle 0, 0, 1 \rangle$ | $\langle 0, 1, 0 \rangle$ | Out-of-plane |
| $d_2$ | 3.49 | $\langle 13.2, 0, 0, 0, 0, 0 \rangle$ | $\langle 1, 0, 0 \rangle$ | $\langle 0, 0.994, 0057 \rangle$ | In-plane |
| $g$ | 2.98 | $\langle 7.21, 7.21, 0, 0, 0, 0.297 \rangle$ | $\langle 0.5, 0.5, 0 \rangle$ | $\langle 0, 0, 1 \rangle$ | In-plane |
| $h$ | 3.19 | $\langle 8.31, 8.31, 0, 0, 0, 0 \rangle$ | $\langle 0.5, 0.5, 0 \rangle$ | $\langle 0.195, 0.038, 0.767 \rangle$ | In-plane |

## 3.2 Polarization identification

As defined previously, wave polarization is the relative orientation between the oscillation direction of the elastic particles and the wave vector. While the wave vector direction is known a priori, the oscillation direction is captured by the modal participation factor, $\mathbf{\Gamma}^n$, which quantifies the contribution of each directional motion component to the mode shape at any given eigenfrequency. Thus, the relative orientation between the two vectors can be calculated as a polarization factor, $\Phi^n$, defined as:

$$\Phi^n = |\cos \varphi^n| \tag{19}$$

where

$$\cos \varphi^n = \frac{\langle \mathbf{\Gamma}_t^n, \mathbf{k}_d \rangle}{\|\mathbf{\Gamma}_t^n\| \|\mathbf{k}_d\|} \tag{20}$$

Here, $\mathbf{\Gamma}_t^n$ and $\mathbf{k}_d$ are the translational components of the modal participation factor and the local reciprocal lattice vector, and are given as:

$$\mathbf{\Gamma}_t^n = \langle \Gamma_1^n, \Gamma_2^n, \Gamma_3^n \rangle \tag{21}$$
$$\mathbf{k}_d = \langle \Delta k_1, \Delta k_2, \Delta k_3 \rangle \tag{22}$$

where, $\Delta k_1, \Delta k_2, \Delta k_3$ are the increment of the wave vector components along the *1*-, *2*- and *3*-axis, respectively.

The polarization factor provides a positive numerical value between 0 and 1, quantifying the relative angle between the wave propagation and lattice motion directions: 0 means that the oscillation is perpendicular to the wave propagation direction—indicating a *S*-wave; 1 means that the oscillation is parallel to the wave propagation direction—indicating a *P*-wave. For cases without pure polarizations, one may use the nomenclature 'quasi-*S*' if the polarization factor is less than 0.1, and 'quasi-*P*' if it is greater than 0.9; waves with polarization values in between these bounds may be classified as 'hybrid' waves.

**Table 2.** The calculated translational modal participation factor, local wave vector orientation, polarization factor, and the identified wave polarization.

| Location | Normalized frequency $\Omega^n$ | Translational components of the modal participation factor $\mathbf{\Gamma}_t^n$ | Local wave vector $\mathbf{k}_d$ | Polarization factor $\Phi^n$ | Wave polarization |
|---|---|---|---|---|---|
| $c_1$ | 3.4128 | $\langle 1.83, 0, 0 \rangle$ | $\langle 1, 0, 0 \rangle$ | 1 | *P*-wave |
| $c_2$ | 3.4037 | $\langle 0, 0, -0.1160 \rangle$ | $\langle 1, 0, 0 \rangle$ | 0 | *SV*-wave |
| $d_1$ | 3.4771 | $\langle 0, 0, -0.15 \rangle$ | $\langle 1, 0, 0 \rangle$ | 0 | *SV*-wave |
| $d_2$ | 3.4954 | $\langle 1.35, 0, 0 \rangle$ | $\langle 1, 0, 0 \rangle$ | 1 | *P*-wave |
| $g$ | 2.9817 | $\langle 0.497, -0.497, 0 \rangle$ | $\langle -1, -1, 0 \rangle$ | 0 | *SH*-wave |
| $h$ | 3.1927 | $\langle 0.566, 0.566, 0 \rangle$ | $\langle -1, -1, 0 \rangle$ | 1 | *P*-wave |

For clarity, reconsider the locations $c_1$, $c_2$, $d_1$, and $d_2$ in Fig. 1 and as discussed above. The calculated modal participation factor, local reciprocal lattice vector, and polarization factor are given in Table 2. The polarization factor at locations $c_1$ and $d_2$ reveals that the oscillation direction of the structure is parallel to the wave propagation direction, thus correctly identifying the eigenfrequency at these locations as a *P*-wave solution. Similarly, the polarization factor at $c_2$ and $d_1$ correctly identifies these locations as an *S*-wave solution; since the dominant motion for both these locations is along the out-of-plane axis, as shown in Table 1, the propagation mode can be further distinguished as an *SV*-wave solution.

The necessity of calculating the polarization factor is reinforced by considering points *g* and *h*—both are at the same wave vectors but at different frequencies. Identifying the polarization of such locations is typically complicated since the wave vector is not oriented along a lattice element. As seen in Table 2, their polarization factors reveal that while oscillation direction at location *g* is perpendicular to the propagating wave direction, the oscillation direction at location *h* is parallel to the wave direction. Thus, the eigenfrequencies at locations *g* and *h* must be classified as *SH*- and *P*-waves, respectively. The calculated modal participation factors, local reciprocal lattice vectors, and polarization factors for all locations shown in Fig. 1 are provided in the Appendix, Table A2.

### *3.3. Visualization Schema*

For easier visualization of the contribution of each relative effective modal mass vector component to the overall motion of the structure at any given eigenfrequency solution, we represent each individual component using the following color scheme:

- $q_{t_1}^n \rightarrow$ solid blue squares
- $q_{t_2}^n \rightarrow$ solid green triangles
- $q_{t_3}^n \rightarrow$ solid red diamonds
- $q_{r_1}^n \rightarrow$ hollow blue circles
- $q_{r_2}^n \rightarrow$ hollow green circles
- $q_{r_3}^n \rightarrow$ hollow red circles

We then overlay these components on the dispersion points and account for each components contribution using the color intensity scales, where for each individual component, the lowest color intensity indicates a numerical value of 0, i.e., the component does not contribute to the motion, and the brightest color intensity indicates a numerical value of 1 and the component's dominance in the overall motion. For clearer visualization, we plot the translational and rotational relative effective modal mass vector components in two separate, adjacent plots. Similarly, to clearly distinguish the *S*- and *P*-waves, we separately plot the polarization factor at each eigenfrequency solution using the asterisk symbol with color intensity varying from cyan to black; cyan indicates a pure *S*-wave and black indicates a pure *P*-wave.

An example application of this visualization scheme is shown in Fig. 2. Here, we plot the dispersion curves of the square lattice previously shown in Fig. 1 using the visualization scheme explained above. As seen from the figures, the proposed visualization scheme makes it easier to accurately identify the wave propagation mode and clearly distinguish the polarization associated with each wave solution. In turn, this makes the identification of the various bandgaps more straightforward and accurate than visually analyzing the mode shapes at each point. For the remainder of the paper, we use this scheme when analyzing the wave propagation behavior of various structures.

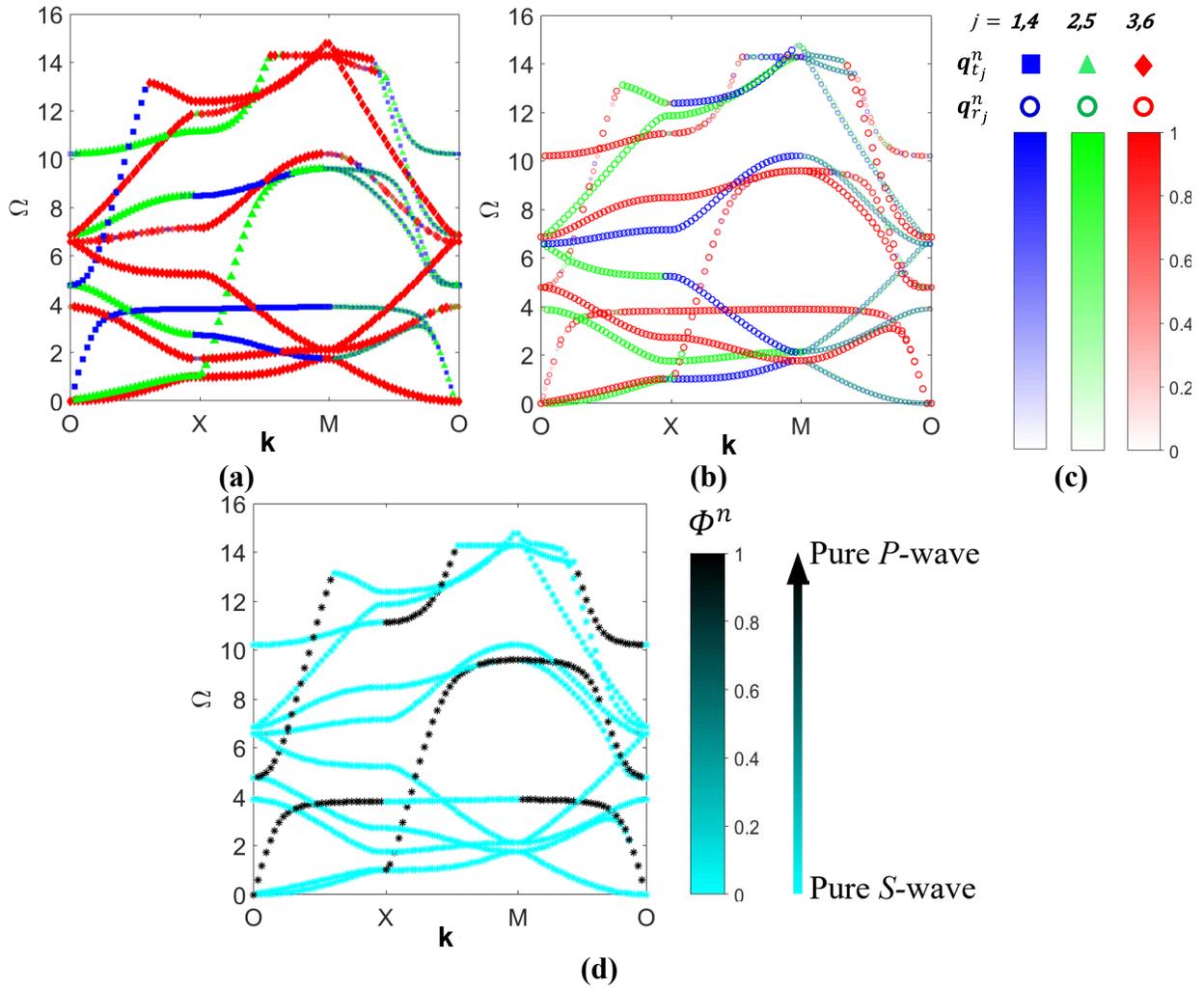

**Fig 2.** Dispersion curves with overlaid relative effective modal mass vector components. **Fig. 2(a)** shows the translational components, **Fig. 2(b)** shows the rotational components, and **Fig. 2(c)** shows the color scales and symbols used for each component. **Fig. 2(d)** shows the polarization plot and the color scale used to identify the different polarization values.

## 4. Method validation

Here, we validate the presented method by comparing the propagation mode and polarization predictions to published results in the literature. First, we verify the method's efficacy in correctly predicting modal coupling by using a coupled beam model. Then, we use the method to identify the presence of wave attenuation bandgaps in a locally resonant sandwich beam, and, finally, a 3D composite metastructure.

### *4.1. Mode identification in a coupled beam*

Consider a prismatic thin-walled cantilevered beam with an open, symmetric channel cross-section, as shown in Fig. 3. Originally studied by Noor et al. [59] using a mixed 1D FE approach and verified using a 2D plate/shell model based on the Sanders-Budiansky shell theory, this beam exhibits strong flexural-torsional coupling that can be difficult to identify without visually analyzing the mode shapes. Here, to demonstrate the accuracy of the proposed method, we analyze the free vibration behavior of the beam and compare the mode types predicted using the relative effective modal mass vectors with those obtained by Noor et al. and by visually inspecting the individual mode shapes. We model the beam as a 3D deformable solid using continuum solid shell elements (CSS8), with its prismatic cross-section in the 2-3 plane and the length along the 1-axis; the geometrical and material parameters are chosen to match those used by Noor et al. and are summarized in Fig. 3.

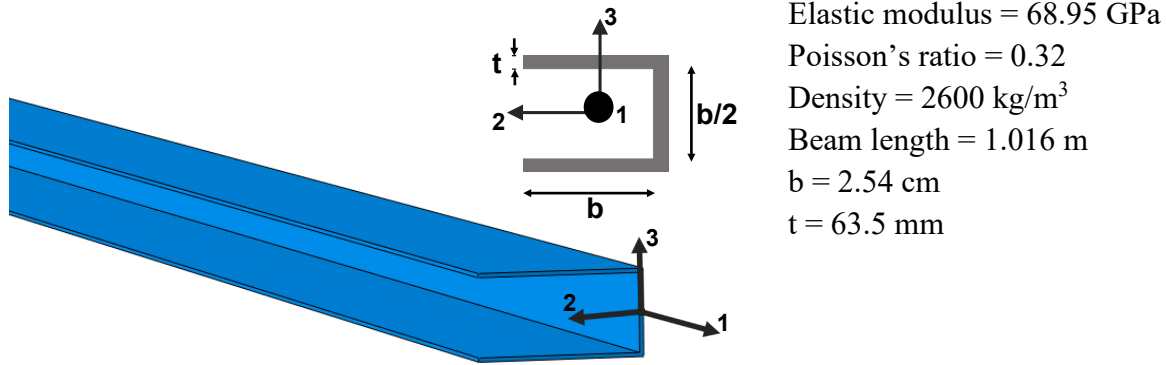

Elastic modulus = 68.95 GPa
Poisson's ratio = 0.32
Density = 2600 kg/m$^3$
Beam length = 1.016 m
b = 2.54 cm
t = 63.5 mm

**Fig. 3.** The analyzed thin-walled cantilever beam with an open, symmetric channel cross-section. The figure shows the geometrical and material properties used by Noor et al. [59].

We use a linear perturbation procedure to extract the first six eigenfrequencies and the corresponding eigenvectors, from which the corresponding relative effective modal mass vectors are calculated to help identify individual modes as flexural, torsional, or coupled (hybrid). The relative effective modal mass components and the mode shapes, as visualized in the solver (Abaqus CAE) are shown for each mode in Table 3.

**Table 3.** The first six natural frequencies, relative effective mass vector components, and the mode shapes obtained from the FE solver.

| Mode number, $n$ | Natural frequency, $\omega^n$ (Hz) | Relative effective translational mass $\mathbf{q}_t^n = \langle q_{t_1}^n, q_{t_2}^n, q_{t_3}^n \rangle$ | Relative effective rotational mass $\mathbf{q}_r^n = \langle q_{r_4}^n, q_{r_5}^n, q_{r_6}^n \rangle$ | Deformed and undeformed mode shapes |
|---|---|---|---|---|
| 1 | 11.43 | $\langle 0, 0, 1 \rangle$ | $\langle 0.372, 0.628, 0 \rangle$ | |
| 2 | 23.158 | $\langle 0, 1, 0 \rangle$ | $\langle 0, 0, 1 \rangle$ | |
| 3 | 42.705 | $\langle 0, 0, 1 \rangle$ | $\langle 0.656, 0.344, 0 \rangle$ | 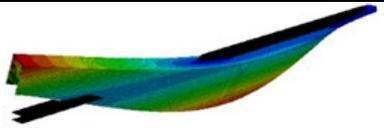 |
| 4 | 57.925 | $\langle 0, 0, 1 \rangle$ | $\langle 0.624, 0.376, 0 \rangle$ | 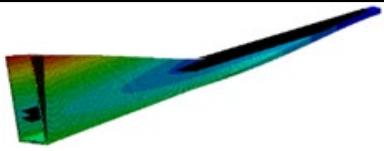 |
| 5 | 106.96 | $\langle 0, 0, 1 \rangle$ | $\langle 0.212, 0.788, 0 \rangle$ | 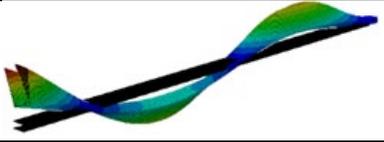 |
| 6 | 144.5 | $\langle 0, 1, 0 \rangle$ | $\langle 0, 0, 1 \rangle$ | 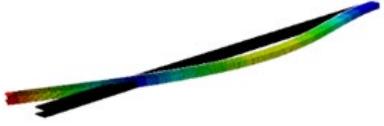 |

In accordance with Noor et al. and the extracted mode shapes, the relative effective modal mass vectors predict strong flexural-torsional coupling in all the modes except the second and sixth modes—these are the first and second horizontal flexural modes, as evidenced by the pure translation along the *2*-axis ($q_{t_2}^n = 1$) and rotation about the *3*-axis ($q_{r_6}^n = 1$). The coupled behavior of the remaining four modes is evidenced by the vertical translation, $q_{t_3}^n$, occurring in combination with the flexure-associated rotation about the *2*-axis, $q_{r_5}^n$, and the torsion-associated rotation about the *1*-axis, $q_{r_1}^n$. Further, the degree of participation of the flexural and torsional motions in each coupled mode is seen by the relative values of the rotations associated with each motion: while the flexure-associated rotation dominates in modes one and five, the torsion-associated rotation dominates in modes three and four. This is also consistent with the strain energy based predictions obtained by Noor et al. Thus, the relative effective modal mass vector accurately identifies all six modes, including the coupled modes.

### *4.2. Propagation mode and bandgap identification in a Timoshenko beam*

Consider a sandwich beam with periodically embedded internal resonators, as studied previously by Sharma et al. [7]. The periodic insertion of spring-mass resonators results in the formation of flexural wave attenuation bandgaps due to local resonance and Bragg scattering effects. Here, we model the sandwich beam as a Timoshenko beam using the material and geometrical parameters as described in Ref. [7] and obtain the dispersion curves using the previously described unit cell approach; the unit cell is chosen such that the resonator is connected to the central node and the Floquet-Bloch boundary conditions are applied at the end nodes. We model the beam as a 3D wire with its length along the *1*-axis, using shear-flexible beam elements with quadratic interpolation (B32). We extract the eigenvalues and eigenvectors using a linear perturbation procedure, carried out while sweeping the wave vector along the longitudinal (*1*-) direction. The extracted eigenvectors are then used to calculate the relative effective modal mass vectors using a Matlab script.

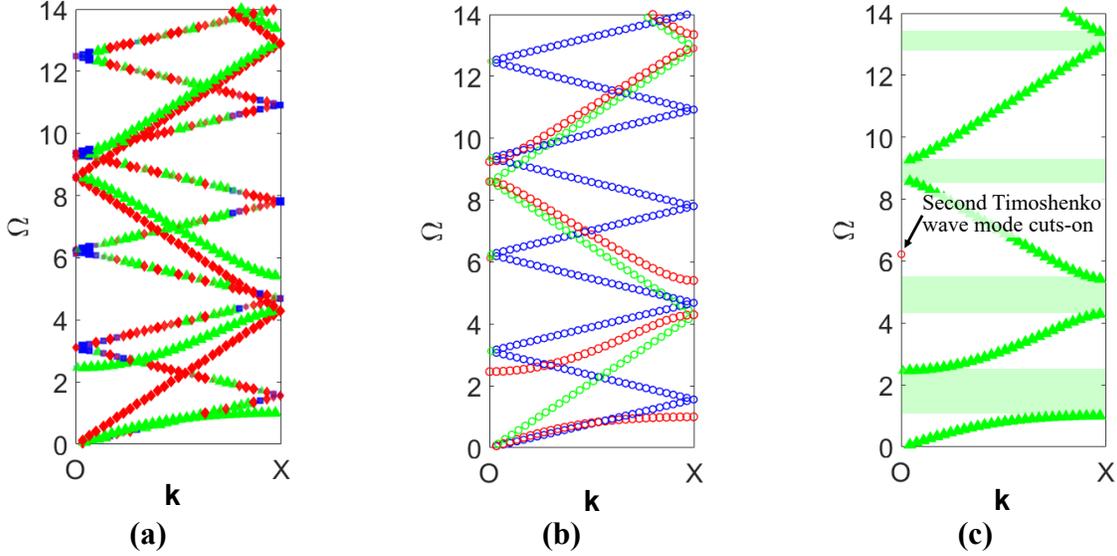

**Fig. 4.** Dispersion curves for the locally resonant sandwich beam, as studied in Ref. [7]. **Fig. 4(a)** shows the translational components and **Fig. 4(b)** shows the rotational components of the relative effective modal mass vectors. **Fig. 4(c)** shows only the flexural wave solution with translational motion along the *2*-direction. The symbols and color scales are as shown in Fig. 2(c).

The dispersion curves with the identified propagation modes for the locally resonant sandwich beam, allowing for motion along all degrees-of-freedom, are shown in Fig. 4(a-b), where all frequencies are normalized with respect to the local resonance frequency (200 Hz). For easier comparison with the published result in [7], Fig. 4(c) shows the dispersion curves obtained when only flexural displacements along the *2*-axis and rotations about the *3*-axis are permitted, in accordance with the assumptions made by Sharma et al. [7]. As expected, the dispersion curves for the sandwich beam with no restrictions on the degrees-of-freedom show multiple wave modes within the frequency range. The periodic insertion of local resonators causes a low-frequency bandgap around the local resonance frequency, and multiple Bragg bandgaps at higher frequencies. Since the resonator motion is restricted to the *2*-direction, all the generated bandgaps are polarized, only attenuating flexural waves with displacements polarized in the *2*-direction. The presence of these polarized flexural bandgaps is seen more clearly in Fig. 4(c), where all the modes except the classical Timoshenko beam modes (flexural and a higher-order mode accounting for the shear deformation and rotational inertia) are suppressed. The predicted bandgaps match exactly the results shown in Fig. 3(b) in [7]. Additionally, the presence of the second Timoshenko mode is accurately captured at $\Omega = 6.12$, and clearly visualized by the hollow red circle, indicating the dominance of rotation about the *3*-axis. Thus, the presented method, coupled with the visualization scheme, allows for an easy identification of each propagating wave mode and, consequently, the accurate assessment of the bandgap polarizations.

## 4.3. Composite 3D-printed metastructure with embedded resonant inclusions

Finally, consider the wave attenuation behavior of the composite 3D-printed metastructure studied by Matlack et al. [21]. The metastructure consists of steel inclusions coated with a thin layer of polycarbonate and periodically embedded within a 3D printed polycarbonate cubic lattice. This design results in the generation of low-frequency attenuation bandgaps driven by the interactions between the local resonance and Bragg bandgaps. Here, we replicate this design using the parameters provided in Ref. [21] and compare the predicted bandgaps with those obtained by Matlack et al. for the high-stiffness case shown in Fig. 2(a) in Ref. [21]. The metastructure is modeled as an infinite beam using three-dimensional, ten-node tetrahedral elements (C3D10). The dispersion curves along the O-X path are then overlaid by the relative modal effective mass vectors and shown in Fig. 5(a-c). For clarity, we plot the displacement components in Fig. 5(a), the rotations in Fig. 5(b), and the polarization factors in Fig. 5(c). We then mark the flexural bandgaps in Fig. 5(a), torsional bandgaps in Fig. 5(b), and longitudinal bandgaps in Fig. 5(c). The accuracy of the bandgap identification—performed without analyzing individual mode shapes for each $\omega(k)$ point—is established by comparing the bandgap predictions in Ref. [21], Fig. S4.

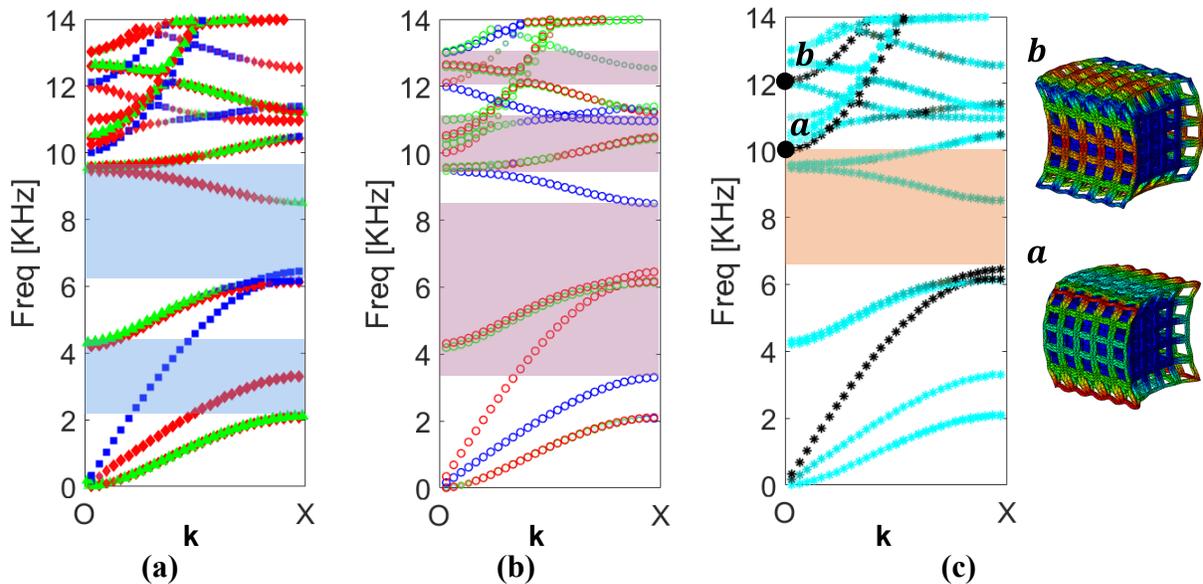

**Fig. 5.** Dispersion curves along the O-X wave vector direction for the composite 3D metastructure studied in Ref. [21]. **Fig. 5(a)** shows the translational components and **Fig. 5(b)** shows the rotational components of the relative effective modal mass vectors. **Fig. 5(c)** shows the polarization factors and the mode shapes at locations *a* and *b*. The shaded regions in Figs. 5(a), (b), and (c) are the identified flexural (blue), torsional (red), and longitudinal (orange) bandgaps.

In agreement with Ref. [21], the high-stiffness composite beam design results in distinct polarized bandgaps, which overlap to form a complete-directional bandgap ranging from 6400 Hz to 8400 Hz. While the predicted flexural bandgaps coincide with those identified in Ref. [21], some additional information is obtained using the relative modal effective mass vectors. First, we identify the presence of two additional torsional bandgaps—as evidenced by the dominant rotation about the axial direction—occurring at higher frequencies; these bandgaps would otherwise be difficult to identify using mode shape visualization. Secondly, the frequency range of the longitudinal bandgap, identified here using the polarization factors, is narrower than that identified in Ref. [21]. Using mode shape observation, the authors incorrectly identify the cut-on frequency of a higher-order longitudinal mode as the bandgap cut-off frequency. In actuality, the longitudinal mode cuts-on at a lower frequency, as correctly predicted by the polarization plot and verified by the mode shapes at locations *a* and *b*, as shown in Fig. 5(c). Thus, the presented method can help avoid the misidentification of propagation modes and bandgaps, especially due to the complex mode shapes occurring at higher frequencies.

## 5. Method Application

Here, as an example application of the developed method, we study the elastic wave propagation behavior of square planar lattices. Specifically, we focus on the existence of polarized bandgaps and the effect of lattice parameter perturbations on the individual bandgaps. We choose the square lattice previously studied in Section 2.1 and shown in Fig. 1 as the base lattice and systematically perturb its side length ratio, vertical strut cross-section, internal angle skewness, and the joint connection type. All the lattices are modeled in Abaqus CAE using B32 beam elements with six degrees-of-freedom—i.e., out of plane modes are permitted—and the dispersion relations and mode identification information are extracted using the proposed method. For consistency of comparison, the mass of all lattices is maintained the same throughout. In the interest of clarity, we discuss the evolution of polarized bandgaps only along the O-X wave vector direction. Here, all eigenfrequencies are normalized with respect to the eigenfrequency of the first in-plane bending mode occurring at high-symmetry point X.

### *5.1. Effect of side length ratio*

We define the side length ratio as $\epsilon = L_2/L_1$, where $L_2$ and $L_1$ are the unit cell lattice constants along the *2*- and *1*-axis, respectively. We modify the square lattice by incrementally increasing $\epsilon$ while maintaining all the other lattice parameters, including its total mass, as constant. The dispersion curves with the identified propagation modes and the associated polarization factors are plotted as a function of $\epsilon$ along the O-X direction in Fig. 6. We plot the translational components in Fig. 6(a), rotational components in Fig. 6(b), and the polarization factors in Fig. 6(c). For clarity, the cut-on and cut-off frequencies for the *SV*-bandgaps are marked in Fig. 6(a) using symbols $v_i$, the *SH*-bandgaps are marked in Fig. 6(b) using symbols $h_i$, and the *P*-bandgaps are marked in Fig. 6(c) using symbols $p_i$. Here, we vary $\epsilon$ from 1 (i.e., square lattice) to $\epsilon = 1.75$; an extreme case of $\epsilon = 3$ is also shown.

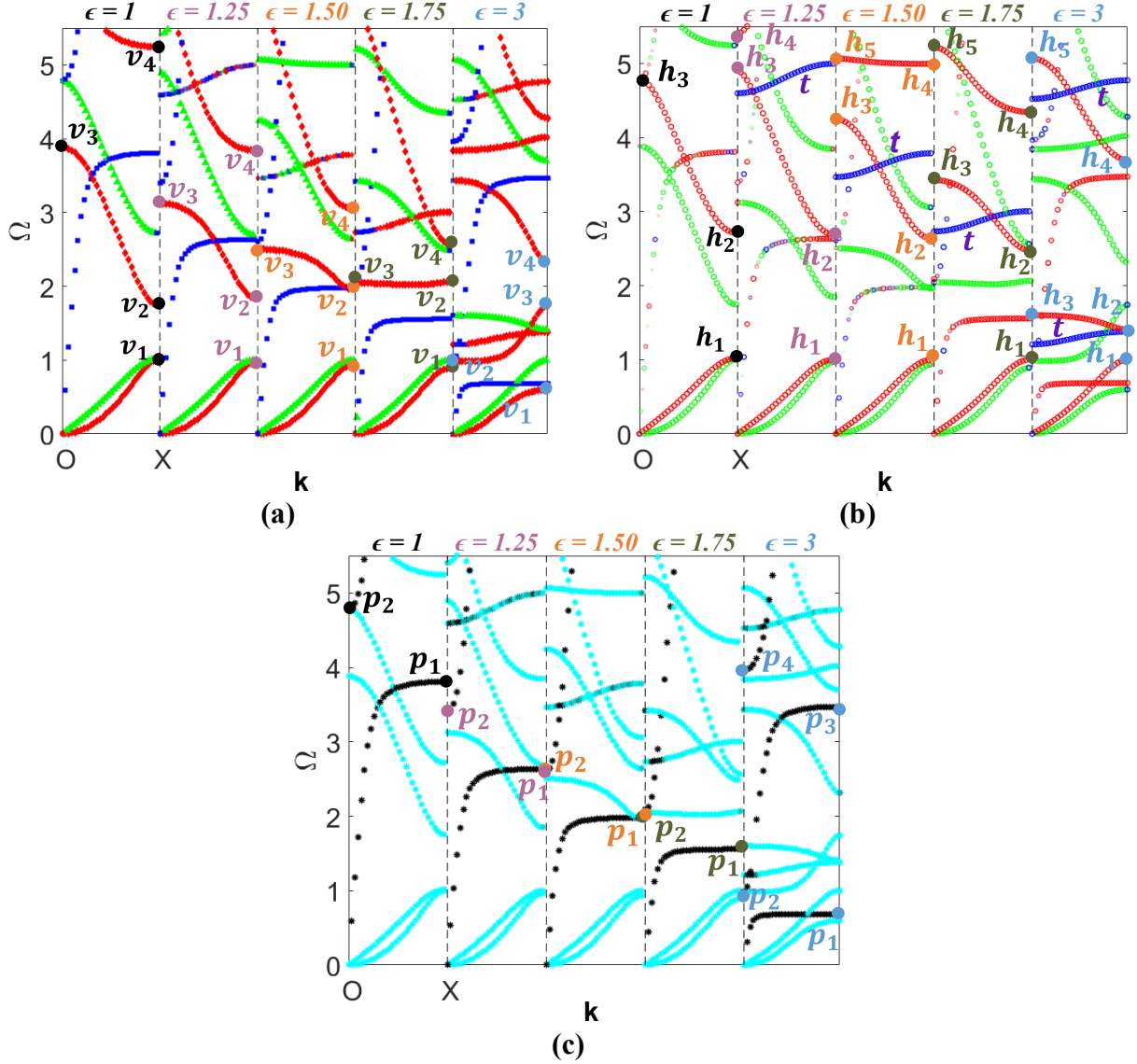

**Fig. 6.** Effect of variation of side length ratio, $\epsilon$, on the dispersion behavior and bandgap evolution along the O-X wave vector direction. The relative effective translational mass and cut-on and cut-off frequencies for *SV*-bandgaps ($v_i$) are marked in **Fig. 6(a)**; the relative effective rotational mass and cut-on and cut-off frequencies for *SH*-bandgaps ($h_i$) are marked in **Fig. 6(b)**; the polarization factors and the cut-on and cut-off frequencies for *P*-bandgaps ($p_i$) are marked in **Fig. 6(c)**. Curves marked as *t* in Fig. 6(b) indicate standing waves with motion dominated by rotations about the *1*-axis.

Along O-X, the dispersion curves for the square lattice, $\epsilon = 1$, show the presence of two *SV*-polarized bandgaps ($v_1$-to-$v_2$ and $v_3$-to-$v_4$), two *SH*-polarized bandgap ($h_1$-to-$h_2$ and $h_3$ and beyond), and one *P*-polarized bandgap ($p_1$-to-$p_2$) within the considered frequency range. The *SV*- and *SH*-bandgaps overlap between $\Omega = 1$ to 1.748, indicating that only *P*-waves can propagate through the lattice along O-X within this frequency range. This behavior is akin to the "fluid-like" behavior recently demonstrated by Ma et al. [22] for a three-dimensional elastic metamaterial with an anisotropic locally resonant unit cell. As the side length ratio is increased, the reduction in lattice symmetry lifts the degeneracy between the *SV*- and *SH*-eigenmodes occurring at the high-symmetry point X at $\Omega = 1$ for the square lattice. As $\epsilon$ increases, the lower bound of the first *SV*-bandgap, $v_1$, gradually shifts to a lower frequency while the upper bound, $v_2$, shifts to a higher frequency, causing an increase in the first *SV*-bandgap width with increasing $\epsilon$. On the other hand, while the lower bound of the first *SH*-bandgap, $h_1$, remains stationary because it is the normalization frequency, the upper bound, $h_2$, gradually shifts to a lower frequency and results in a reduction in the first *SH*-bandgap width with increasing $\epsilon$. While these changes in the first *SH*- and *SV*-bandgap widths may be considered insignificant, the changes observed in the higher order bandgaps are more pronounced. Increasing $\epsilon$ reduces the width of the second *SV*-bandgap while shifting it to a comparatively lower frequency range. Similarly, the second *SH*-bandgap shifts to lower frequencies with increasing $\epsilon$; however, in contrast with the second *SV*-bandgap, the width of this bandgap increases with increasing $\epsilon$ and an ultrawide *SH*-bandgap emerges for the extreme case of $\epsilon = 3$. Further, the comparative beginning and end locations of the bandgap bounding curves indicate that for both *SH*- and *SV*-polarizations, the first bandgaps are driven by Bragg effects while the second bandgaps occur due to lattice resonance effects [7]. These observations align with the fact that increasing $\epsilon$ does not alter the lattice periodicity along O-X, but it reduces the effective lattice stiffness along the *2*- and *3*-axis, thus reducing the lattice structural resonance frequencies responsible for the second bandgaps. For the extreme case of $\epsilon = 3$, the first *SV*-bandgap changes into a resonance-driven gap while the second *SV*-bandgap changes into a Bragg bandgap. Interestingly, a nearly flat band, marked as *t* in Fig. 6(b), with dominant rotations about the *1*-axis appears at $\epsilon = 1.25$ and shifts to lower frequencies with increasing $\epsilon$; two such bands are observed for $\epsilon = 3$. The flatness of these bands indicate that these are rotational standing waves resembling torsional behavior in finite structures.

The effect of this stiffness reduction is also observed on the *P*-bandgap, which is generated due to the first structural resonance occurring in the vertical lattice struts oriented perpendicular to the wave propagation direction, as evidenced by the behavior of the upper and lower bounding curves. As $\epsilon$ increases, the reduction in the effective stiffness reduces the resonance frequency, $p_1$, and shifts the *P*-bandgap to a lower frequency. Eventually, for the case of $\epsilon = 1.75$, the *P*-bandgap overlaps with the first *SV*- and *SH*-bandgaps between $\Omega = 1.75$ to 2.5, resulting in the emergence of a complete-directional bandgap where all waves along O-X are spatially attenuated, irrespective of their polarizations. For the extreme case of $\epsilon = 3$, the interactions between the individual resonance and Bragg effects result in the uncoupling of the first *SV*- and *SH*-bandgaps, but an overlap of the *P*- and *SV*-bandgaps results in the emergence of a frequency region where only waves with *SH*-polarization can propagate through the lattice.

### *5.2. Effect of lattice internal angle*

The variation in the dispersion curves and the wave polarizations along the O-X direction as a function of the lattice internal angle, $\Theta$, are shown in Fig 7. Here, we vary $\Theta$ by changing the orientation of the vertical strut with respect to the horizontal lattice strut. In contrast to the variation in $\epsilon$, a change in $\Theta$ drastically changes the dispersion behavior of the lattice, indicating that the elastic wave propagation behavior of the lattice is significantly more sensitive to rotational symmetries in the lattice. Given the high number of low frequency modes for the skewed lattices, we restrict the analysis in this case to $\Omega = 2.5$.

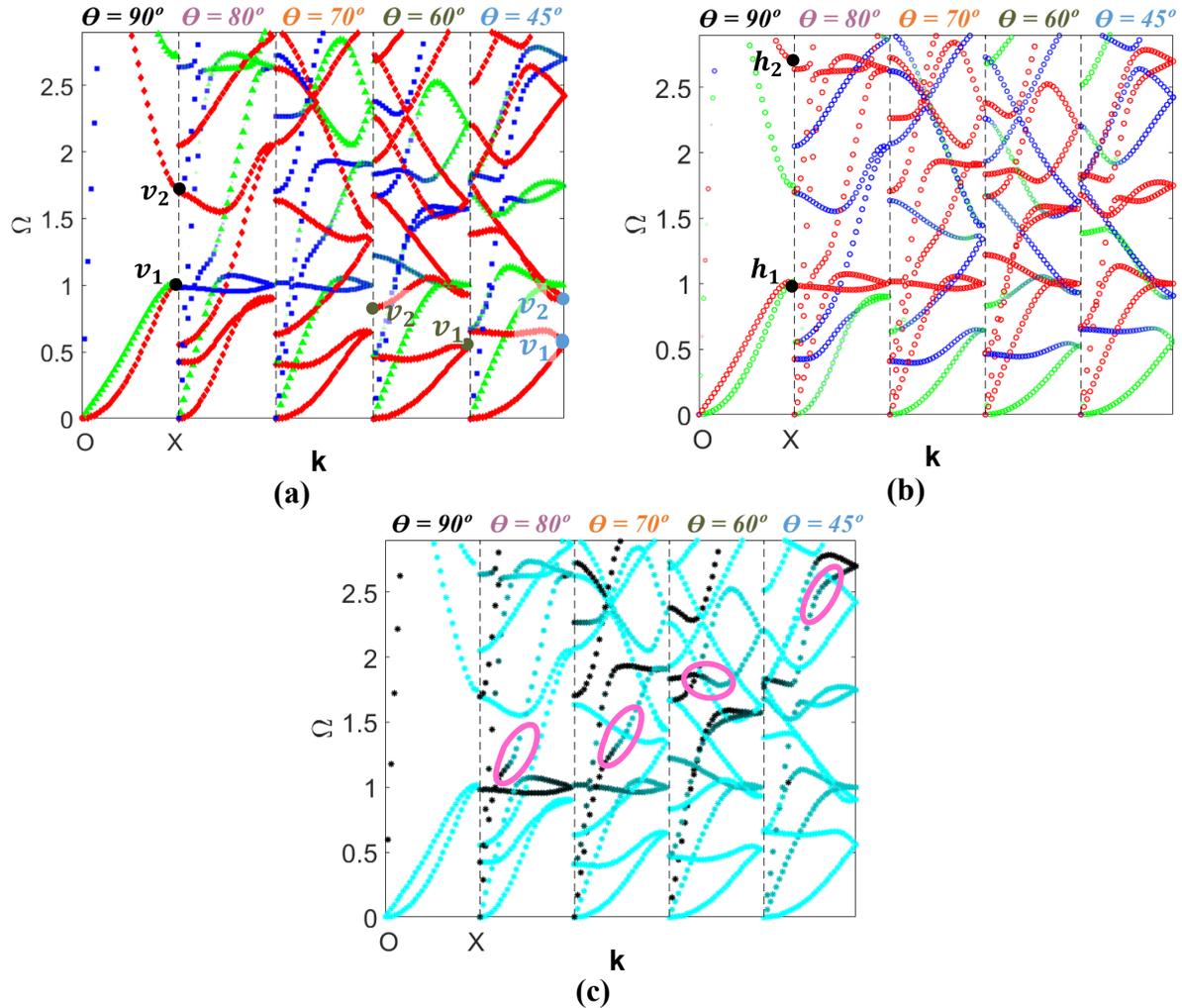

**Fig. 7.** Effect of variation of lattice internal angle, $\Theta$, on the dispersion behavior and bandgap evolution along the O-X wave vector direction. The relative effective translational mass and cut-on and cut-off frequencies for *SV*-bandgaps ($v_i$) are shown in **Fig. 7(a)**; the relative effective rotational mass and cut-on and cut-off frequencies for *SH*-bandgaps ($h_i$) are shown in **Fig. 7(b)**; the polarization factors are shown in **Fig. 7(c)**, where the elliptical markings show examples of mode conversion between *P*- and *S*-wave propagation modes.

Altering the square lattice ($\Theta = 90º$) to a skewed lattice with $\Theta = 80º$ results in the disappearance of the previously observed Bragg *SV*- and *SH*-bandgaps originating at $\Omega = 1$, marked as $v_1$-to-$v_2$ and $h_1$-to-$h_2$, respectively. Instead, the angular distortion results in the appearance of multiple new *SV*-propagation modes, distinct from those observed within the same frequency range for the square lattice—the new modes are dominated by out-of-plane motion coupled with torsion-like rotations about the *1*-axis. As the angular distortion is increased, a low-frequency *SV*-bandgap reappears for the cases of = 60º and 45º. Altering the orientation of the vertical struts also introduces additional stiffness in the horizontal direction, as evidenced by the increase in the slope of the first *SH*-curve. Further, the relative orientation of the lattice struts with respect to the wave vector orientation causes in-plane coupling and mode transitions in the *P*- and *SH*-modes. These coupled in-plane propagation modes are more clearly observed in the polarization angle plots, Fig. 7(c), where the darker dispersion points indicate a higher contribution of lattice motion parallel to the wave vector direction—representative cases of transitions between the two modes within the same propagation band are marked in Fig. 7(c). Thus, the polarization plots allow an easier identification of the *P*-wave solutions and the corresponding identification of frequency regions where such waves cannot propagate.

### *5.3. Effect of vertical strut cross-section width*

In this case, we modify the lattice by changing the cross-sectional shape of the vertical strut; the vertical strut is altered by changing the ratio of the cross-sectional widths along the *1*- and the *3*-direction. Taking this cross-sectional width ratio equal to 1 as the base lattice configuration—i.e., a square lattice made using square horizontal and vertical struts—we alter the vertical strut shape by increasing its cross-sectional width ratio while maintaining its cross-sectional area, and consequently the overall lattice mass, as a constant. We use the symbol *V* to denote the rectangularity of the vertical strut. All other lattice parameters, including the lattice constants $L_1$ and $L_2$ are kept constant.

The dispersion curves and the wave polarizations along the O-X direction as a function of the *V*, are shown in Fig. 8. Overall, the first *SV*-bandgap remains unaffected by the changes in *V* since this bandgap is generated due to the Bragg effects and is bounded by the out-of-plane motion of the horizontal struts. While the lower bounding frequency of the first *SH*-bandgap ($h_1$) is similarly unaffected by changes in *V*, the upper bounding frequency ($h_2$) increases with increasing *V*, resulting in widening this bandgap. By contrast, the second *SV*-bandgap ($v_3$-to-$v_4$), generated by structural resonance effects, shifts to a lower frequency with increasing *V* because of the accompanying reduction in the out-of-plane stiffness of the vertical struts. This gap eventually overlaps with the first *SH*-bandgap and forms a directional *S*-bandgap where only *P*-waves can propagate (*V* = 2.5). Note that a similar bandgap was observed when increasing $\epsilon$ in section 5.1; however, in that case, the overlapping *SV*- and *SH*-bandgaps are both generated by Bragg effects. Further, increasing V increases the vertical strut's in-plane stiffness and causes the first *P*-bandgap ($p_1$-to-$p_2$)—driven by vertical strut resonances—to shift to higher frequencies. A propagation mode dominated by rotations about the *1*-axis, marked as *t* in Fig. 8(b) and similar to that observed in section 5.1, is observed around $\Omega$ = 5 for *V* = 2. This rotational mode with significant coupling between the *P*- and *SV*-modes, likely resulting from the vertical strut cross-sectional asymmetry, shifts to lower frequencies with increasing *V*, eventually converting into a propagation mode that transitions from a pure *P*-mode to a pure *SV*-mode for the extreme case of *V* = 4.

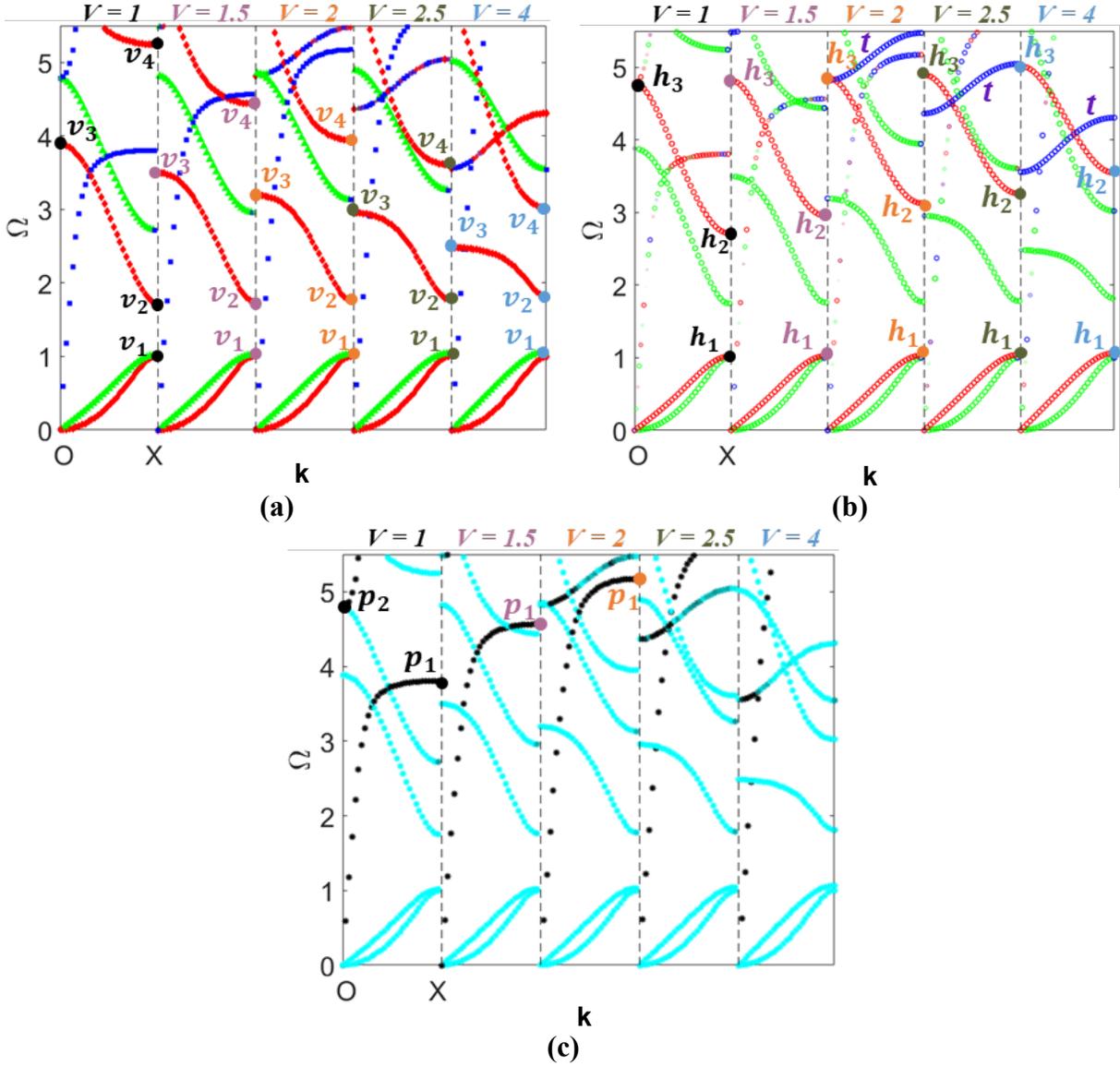

**Fig. 8.** Effect of variation of vertical strut cross-sectional width ratio, $V$, on the dispersion behavior and bandgap evolution along the O-X wave vector direction. The relative effective translational mass and cut-on and cut-off frequencies for *SV*-bandgaps ($v_i$) are shown in **Fig. 8(a)**; the relative effective rotational mass and cut-on and cut-off frequencies for *SH*-bandgaps ($h_i$) are shown in **Fig. 8(b)**; the polarization factors and the cut-on and cut-off frequencies for *P*-bandgaps ($p_i$) are marked in **Fig. 8(c)**. Curves marked as $t$ in Fig. 8(b) indicate standing waves with motion dominated by rotations about the *1*-axis.

## *5.4 Effect of change in lattice joint*

Here, we consider the effect of changing the nature of the joint connecting the horizontal and vertical struts of a square lattice. Previously, Wang et al. [60] have considered the effect of joint connectivity on the wave propagation behavior of triangular and hexagonal Euler-Bernoulli lattice structures. Their results demonstrate that altering the connectivity induces local resonance effects which may or may not result in the generation of complete bandgaps, depending on the global lattice topology.

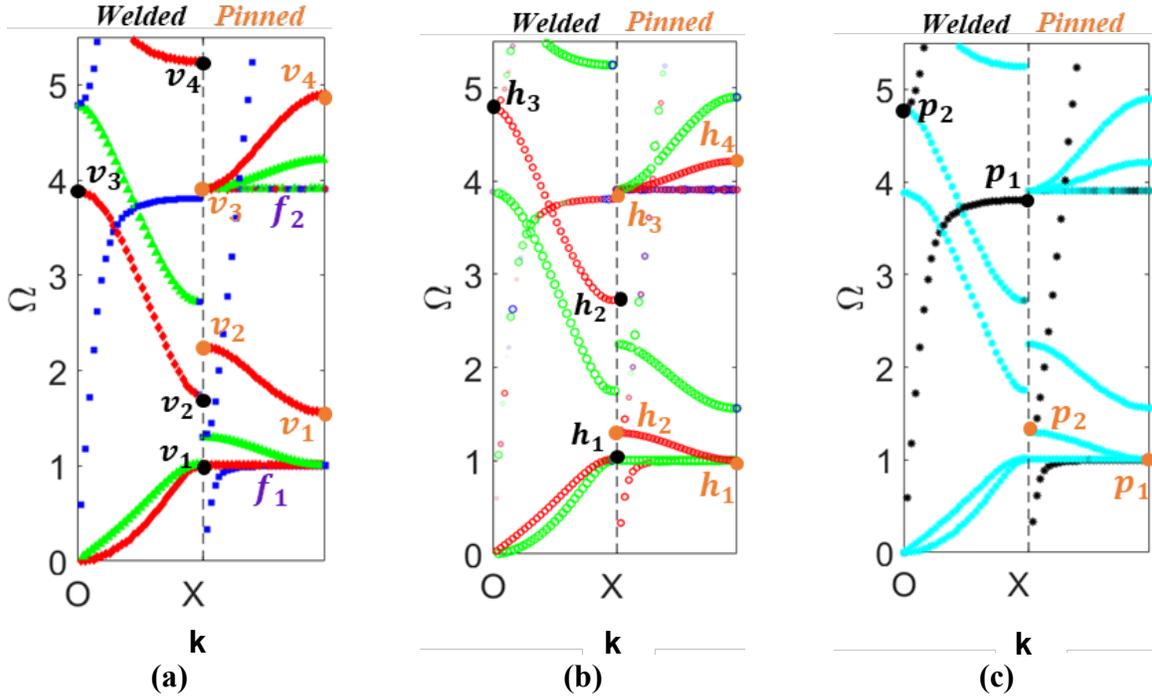

**Fig. 9.** Effect of changing lattice joint type from welded to pinned on the dispersion behavior and bandgap evolution along the O-X wave vector direction. The relative effective translational mass and cut-on and cut-off frequencies for *SV*-bandgaps ($v_i$) are shown in **Fig. 9(a)**; the relative effective rotational mass and cut-on and cut-off frequencies for *SH*-bandgaps ($h_i$) are shown in **Fig. 9(b)**; the polarization factors and the cut-on and cut-off frequencies for *P*-bandgaps ($p_i$) are marked in **Fig. 9(c)**. Curves marked as $f_1$ and $f_2$ in Fig. 9(a) indicate standing waves or structural resonances in the pinned joint lattice.

The dispersion curves and wave polarization plots of the square lattices with welded and pinned joints are compared in Fig. 9. In line with the observations by Wang et al. [60], altering the joint connectivity of the square lattice to a pinned joint significantly changes its dispersion behavior; specifically, two flat dispersion curves (marked as $f_1$ and $f_2$)—indicating standing waves or structural resonances—absent in the welded lattice are observed for the pinned lattice. Interestingly, the pinned lattice shows multiple, wide *S*-bandgaps, including ultralow-frequency *SH*- and *SV*-bandgaps that extend to 0 Hz—indicating a propagation cut-on frequency below which all *S*-waves are spatially attenuated. The cut-on frequency for *SH*-waves ($h_1$) is lower than that for the *SV*-waves ($v_1$), i.e., the pinned lattice predominantly behaves as a fluid-like structure by only supporting the propagation of *P*-waves below the first lattice element flexural resonance frequency. A *P*-bandgap ($p_1$-to-$p_2$) is also observed above this resonance frequency, which overlaps with the *SV*-bandgap and results in a frequency zone where only *SH*-waves can propagate through the structure. The second *SH*- and *SV*-bandgaps are caused due to Bragg effects, and both terminate at the second structural resonance ($v_3$ and $h_3$ are located on the flat band $f_2$). The overlap between these bandgaps generates a second fluid-like behavior frequency region extending between $v_2$ and $v_3$. It should be noted that the presence of an ultralow *S*- bandgap can also be seen in the results by Wang et al. (see Fig. S8(d) in Supplementary Information accompanying [60]); the presented visualization scheme makes it easier to identify this and other emergent behaviors, such as fluid-like behavior of square lattices with pinned joints, which may otherwise be easily missed.

## 6. Conclusion

In this paper, we presented a new method for accurately identifying the propagation mode and polarization of elastic waves traveling in periodic structures and architected metamaterials. The method, proposed as an alternative to the commonly used visual mode inspection technique, adapts the concept of modal participation factor to quantify the contribution of each translational and rotational motion component to the overall wave motion of the structure at each dispersion solution. While this information is sufficient for identifying the associated polarization of low-frequency elastic waves in structures oriented along the wave vector, we propose the use of a new polarization factor to avoid misidentification of polarizations for more complex cases. The robustness of the proposed method was demonstrated by comparing our predictions against previously presented results. Finally, we apply the developed method to study the effect of lattice and structural parameter perturbations on the wave propagation behavior of a planar square beam lattice. We prove the effectiveness of the method by demonstrating the emergence of various previously unobserved directional and polarized bandgaps and novel dynamic properties, such fluid-like behavior and ultralow-frequency bandgaps. Thus, the presented method provides a robust computational approach for studying periodic media with complex elastic wave propagation behavior.

**Acknowledgement**

This research did not receive any specific grant from funding agencies in the public, commercial, or not-for-profit sectors.

**Appendix**

**Table A1.** The calculated modal effective mass vectors, relative effective modal mass vectors, and the identified propagation mode for the locations marked in Fig. 1.

| Location | Normalized frequency $\Omega^n$ | Modal effective mass vectors $\mathbf{m}_{\text{eff}}^n$ | Relative effective translational mass $\mathbf{q}_t^n = \langle q_{t_1}^n, q_{t_2}^n, q_{t_3}^n \rangle$ | Relative effective rotational mass $\mathbf{q}_r^n = \langle q_{r_4}^n, q_{r_5}^n, q_{r_6}^n \rangle$ | Propagation mode |
|---|---|---|---|---|---|
| a | 0.978 | $\langle 0, 0, 20.931, 0, 0.010, 0 \rangle$ | $\langle 0, 0, 1 \rangle$ | $\langle 0, 1, 0 \rangle$ | *Out-of-plane* |
| b | 1.000 | $\langle 0.0, 21.202, 0, 0, 0, 0.002 \rangle$ | $\langle 0, 1, 0 \rangle$ | $\langle 0, 0, 1 \rangle$ | *In-plane* |
| c | 1.738 | $\langle 0, 0, 0.146, 0, 0.985, 0 \rangle$ | $\langle 0, 0, 1 \rangle$ | $\langle 0, 1, 0 \rangle$ | *Out-of-plane* |
| $c_1$ | 3.412 | $\langle 14.390, 0, 0, 0, 0, 0 \rangle$ | $\langle 1, 0, 0 \rangle$ | $\langle 0.003, 0.996, 0 \rangle$ | *In-plane* |
| $c_2$ | 3.402 | ⟨0, 0, 0.183, 0, 0.292, 0⟩ | ⟨0, 0, 1⟩ | ⟨0, 1, 0⟩ | *Out-of-plane* |
| $d_1$ | 3.481 | ⟨0, 0, 0.125, 0, 0.246, 0⟩ | ⟨0, 0, 1⟩ | ⟨0, 1, 0⟩ | *Out-of-plane* |
| $d_2$ | 3.493 | ⟨13.239, 0, 0, 0, 0, 0⟩ | ⟨1, 0, 0⟩ | ⟨0, 0.994, 006⟩ | *In-plane* |
| $e_1$ | 10.079 | ⟨0, 0.02, 0, 0, 0, 0⟩ | ⟨0, 1, 0⟩ | ⟨0, 0, 1⟩ | *In-plane* |
| $e_2$ | 10.985 | ⟨0, 1.233, 0, 0, 0, 0⟩ | ⟨0, 1, 0⟩ | ⟨0, 0, 1⟩ | *In-plane* |
| $f_1$ | 2.692 | ⟨0, 0.013, 0, 0, 0, 1.260⟩ | ⟨0, 1, 0⟩ | ⟨0, 0, 1⟩ | *In-plane* |
| $f_2$ | 1.866 | ⟨14.4, 0, 0, 0, 0, 0⟩ | ⟨1, 0, 0⟩ | ⟨0, 0, 1⟩ | *In-plane* |
| $g$ | 2.985 | ⟨7.21, 7.21, 0, 0, 0, 0.297⟩ | ⟨0.5, 0.5, 0⟩ | ⟨0, 0, 1⟩ | *In-plane* |
| $h$ | 3.193 | ⟨8.309, 8.309, 0, 0, 0, 0⟩ | ⟨0.5, 0.5, 0⟩ | ⟨0.195, 0.038, 0.767⟩ | *In-plane* |

**Table A2.** The calculated translational modal participation factor, local wave vector orientation, polarization factor, and the identified wave polarization for the locations marked in Fig. 1.

| Location | Normalized frequency $\Omega^n$ | Translational components of the modal participation factor $\mathbf{\Gamma}_t^n$ | Local wave vector $\mathbf{k}_d$ | Polarization factor $\Phi^n$ | Wave polarization |
|---|---|---|---|---|---|
| $a$ | 0.978 | ⟨0, 0, 0.691⟩ | ⟨1, 0, 0⟩ | 0 | *SV*-wave |
| $b$ | 1.00 | ⟨0, 0.976, 0⟩ | ⟨1, 0, 0⟩ | 0 | *SH*-wave |
| $c$ | 1.738 | ⟨0, 0, 0.143⟩ | ⟨1, 0, 0⟩ | 0 | *SV*-wave |
| $c_1$ | 3.412 | ⟨1.827, 0, 0⟩ | ⟨1, 0, 0⟩ | 1 | *P*-wave |
| $c_2$ | 3.402 | ⟨0, 0, −0.116⟩ | ⟨1, 0, 0⟩ | 0 | *SV*-wave |
| $d_1$ | 3.481 | ⟨0, 0, −0.150⟩ | ⟨1, 0, 0⟩ | 0 | *SV*-wave |
| $d_2$ | 3.493 | ⟨1.346, 0, 0⟩ | ⟨1, 0, 0⟩ | 1 | *P*-wave |
| $e_1$ | 10.079 | ⟨0, −0.013, 0⟩ | ⟨1, 0, 0⟩ | 0 | *SH*-wave |
| $e_2$ | 10.985 | ⟨0, 0.493, 0⟩ | ⟨1, 0, 0⟩ | 0 | *SH*-wave |
| $f_1$ | 2.692 | ⟨0, 0.007, 0⟩ | ⟨1, 0, 0⟩ | 0 | *SH*-wave |
| $f_2$ | 1.866 | ⟨0.156, 0, 0⟩ | ⟨0, 1, 0⟩ | 0 | *SH*-wave |
| $g$ | 2.985 | ⟨0.497, −0.497, 0⟩ | ⟨−1, −1, 0⟩ | 0 | *SH*-wave |
| $h$ | 3.193 | ⟨0.566, 0.566, 0⟩ | ⟨−1, −1, 0⟩ | 1 | *P*-wave |

**References**

[1] M.I. Hussein, M.J. Leamy, M. Ruzzene, Dynamics of Phononic Materials and Structures: Historical Origins, Recent Progress, and Future Outlook, Applied Mechanics Reviews, 66 (2014).
[2] S.A. Cummer, J. Christensen, A. Alù, Controlling sound with acoustic metamaterials, Nature Reviews Materials, 1 (2016) 1-13.
[3] G. Ma, P. Sheng, Acoustic metamaterials: From local resonances to broad horizons, Science advances, 2 (2016) e1501595.
[4] L. Brillouin, Wave propagation in periodic structures, 2nd Edition, Dover, New York (1946).
[5] T.J. Cui, D.R. Smith, R. Liu, Metamaterials, Springer, 2010.

[6] L. Liu, M.I. Hussein, Wave motion in periodic flexural beams and characterization of the transition between Bragg scattering and local resonance, Journal of Applied Mechanics, 79 (2012).
[7] B. Sharma, C.-T. Sun, Local resonance and Bragg bandgaps in sandwich beams containing periodically inserted resonators, Journal of Sound and Vibration, 364 (2016) 133-146.
[8] L. Raghavan, A.S. Phani, Local resonance bandgaps in periodic media: Theory and experiment, The Journal of the Acoustical Society of America, 134 (2013) 1950-1959.
[9] J. Achenbach, Wave propagation in elastic solids, Elsevier, 2012.
[10] J.F. Doyle, Wave propagation in structures, in: Wave Propagation in Structures, Springer, 1989, pp. 126-156.
[11] G. Armstrong, Coupling between torsional and bending modes of vibration in cantilever beams, in, Durham University, 1972.
[12] B.R. Mace, E. Manconi, Wave motion and dispersion phenomena: Veering, locking and strong coupling effects, The Journal of the Acoustical Society of America, 131 (2012) 1015-1028.
[13] E. Manconi, B. Mace, Veering and Strong Coupling Effects in Structural Dynamics, Journal of Vibration and Acoustics, 139 (2017) 021009.
[14] S. Mizuno, Resonance and mode conversion of phonons scattered by superlattices with inhomogeneity, Physical Review B, 68 (2003) 193305.
[15] N. Joel, Reflection and Polarization of Elastic Waves in a LiF Crystal: Mode Conversion from Longitudinal to Transverse, Proceedings of the Physical Society (1958-1967), 78 (1961) 38.
[16] J. Willis, Polarization approach to the scattering of elastic waves—I. Scattering by a single inclusion, Journal of the Mechanics and Physics of Solids, 28 (1980) 287-305.
[17] B. Manzanares-Martinez, F. Ramos-Mendieta, Sagittal acoustic waves in phononic crystals: k-dependent polarization, Physical Review B, 76 (2007) 134303.
[18] D.L. Anderson, Elastic wave propagation in layered anisotropic media, Journal of Geophysical Research, 66 (1961) 2953-2963.
[19] M. Musgrave, The propagation of elastic waves in crystals and other anisotropic media, Reports on progress in physics, 22 (1959) 74.
[20] A. Bacigalupo, M. Lepidi, Acoustic wave polarization and energy flow in periodic beam lattice materials, International Journal of Solids and Structures, 147 (2018) 183-203.
[21] K.H. Matlack, A. Bauhofer, S. Krodel, A. Palermo, C. Daraio, Composite 3D-printed metastructures for low-frequency and broadband vibration absorption, Proc Natl Acad Sci U S A, 113 (2016) 8386-8390.
[22] G. Ma, C. Fu, G. Wang, P. Del Hougne, J. Christensen, Y. Lai, P. Sheng, Polarization bandgaps and fluid-like elasticity in fully solid elastic metamaterials, Nature communications, 7 (2016) 1-8.
[23] A. Bayat, S. Gaitanaros, Wave Directionality in Three-Dimensional Periodic Lattices, Journal of Applied Mechanics, 85 (2017).
[24] M. Ruzzene, F. Scarpa, F. Soranna, Wave beaming effects in two-dimensional cellular structures, Smart Materials and Structures, 12 (2003) 363-372.

[25] A.J. Zelhofer, D.M. Kochmann, On acoustic wave beaming in two-dimensional structural lattices, International Journal of Solids and Structures, 115-116 (2017) 248-269.

[26] B. Manzanares-Martínez, J. Sánchez-Dehesa, A. Håkansson, F. Cervera, F. Ramos-Mendieta, Experimental evidence of omnidirectional elastic bandgap in finite one-dimensional phononic systems, Applied physics letters, 85 (2004) 154-156.

[27] Y. Wang, Z. Li, M.V. Golub, G. Huang, W. Chen, C. Zhang, Interfacial delamination-induced unidirectional propagation of guided waves in multilayered media, Mathematics and Mechanics of Solids, (2022) 10812865221092680.

[28] H. Miao, F. Li, Shear horizontal wave transducers for structural health monitoring and nondestructive testing: A review, Ultrasonics, 114 (2021) 106355.

[29] K. Helbig, J.M. Carcione, Anomalous polarization in anisotropic media, European Journal of Mechanics-A/Solids, 28 (2009) 704-711.

[30] H.J. Lee, J.R. Lee, S.H. Moon, T.J. Je, E.C. Jeon, K. Kim, Y.Y. Kim, Off-centered Double-slit Metamaterial for Elastic Wave Polarization Anomaly, Sci Rep, 7 (2017) 15378.

[31] G.U. Patil, K.H. Matlack, 3D auxetic lattice materials for anomalous elastic wave polarization, J Acoust Soc Am, 145 (2019) 1259.

[32] H. Huang, C. Sun, Wave attenuation mechanism in an acoustic metamaterial with negative effective mass density, New Journal of Physics, 11 (2009) 013003.

[33] D. Mead, A general theory of harmonic wave propagation in linear periodic systems with multiple coupling, Journal of Sound and Vibration, 27 (1973) 235-260.

[34] D. Mead, S. Parthan, Free wave propagation in two-dimensional periodic plates, Journal of Sound and Vibration, 64 (1979) 325-348.

[35] D. Mead, K. Pujara, Space-harmonic analysis of periodically supported beams: response to convected random loading, Journal of sound and vibration, 14 (1971) 525-541.

[36] D.J. Mead, Free wave propagation in periodically supported, infinite beams, Journal of Sound and Vibration, 11 (1970) 181-197.

[37] D.J. Mead, Wave propagation and natural modes in periodic systems: I. Mono-coupled systems, Journal of Sound and Vibration, 40 (1975) 1-18.

[38] Y.-K. Lin, T. McDaniel, Dynamics of beam-type periodic structures, (1969).

[39] D.J. Mead, A new method of analyzing wave propagation in periodic structures; applications to periodic Timoshenko beams and stiffened plates, Journal of Sound and Vibration, 104 (1986) 9-27.

[40] R.M. Orris, M. Petyt, A finite element study of harmonic wave propagation in periodic structures, Journal of Sound and Vibration, 33 (1974) 223-236.

[41] W. Axmann, P. Kuchment, An efficient finite element method for computing spectra of photonic and acoustic band-gap materials: I. Scalar case, Journal of Computational Physics, 150 (1999) 468-481.

[42] M. Åberg, P. Gudmundson, The usage of standard finite element codes for computation of dispersion relations in materials with periodic microstructure, The Journal of the Acoustical Society of America, 102 (1997) 2007-2013.

[43] N.G. Guarin-Zapata, J., Evaluation of the Spectral Finite Element Method With the Theory of Phononic Crystals, Journal of Computational Acoustics, 23 (2015).
[44] J.F. Doyle, Nonlinear structural dynamics using FE methods, Cambridge University Press, 2014.
[45] C. Kittel, P. McEuen, P. McEuen, Introduction to solid state physics, Wiley New York, 1996.
[46] A.S. Phani, M.I. Hussein, Dynamics of lattice materials, John Wiley & Sons, 2017.
[47] Y. Achaoui, A. Khelif, S. Benchabane, V. Laude, Polarization state and level repulsion in two-dimensional phononic crystals and waveguides in the presence of material anisotropy, Journal of Physics D: Applied Physics, 43 (2010) 185401.
[48] I.J. Pérez-Arriaga, G.C. Verghese, F.C. Schweppe, Selective modal analysis with applications to electric power systems, Part I: Heuristic introduction, ieee transactions on power apparatus and systems, (1982) 3117-3125.
[49] P. Jayakumar, Modeling and identification in structural dynamics, (1987).
[50] W.T. Thomson, Theory of vibration with applications, CrC Press, 2018.
[51] M. Ruzzene, F. Scarpa, F. Soranna, Wave beaming effects in two-dimensional cellular structures, Smart Materials and Structures, 12 (2003).
[52] A.S. Phani, J. Woodhouse, N.A. Fleck, Wave propagation in two-dimensional periodic lattices, The Journal of the Acoustical Society of America, 119 (2006) 1995-2005.
[53] N. Perkins, C. Mote Jr, Comments on curve veering in eigenvalue problems, Journal of Sound and Vibration, 106 (1986) 451-463.
[54] A.S. Veletsos, and Carlos E. Ventura, Modal analysis of non-classically damped linear systems, Earthquake engineering & structural dynamics, 14.2 (1986) 217-243.
[55] J.-T. Chen, H-K. Hong, and Chau-Shioung Yeh, Modal reaction method for modal participation factors in support motion problems, Communications in numerical methods in engineering, 11.6 (1995) 479-490.
[56] U. Füllekrug, Determination of effective masses and modal masses from base-driven tests, International Modal Analysis Conference(IMAC), 14 th, Dearborn, MI., (1996).
[57] M. Nieto, M. Elsayed, D. Walch, Modal participation factors and their potential applications in aerospace: A Review, (2018).
[58] M. Aenlle, M. Juul, R. Brincker, Modal mass and length of mode shapes in structural dynamics, Shock and Vibration, 2020 (2020).
[59] A.K. Noor, J.M. Peters, B.-J. Min, Mixed finite element models for free vibrations of thin-walled beams, Finite elements in analysis and design, 5 (1989) 291-305.
[60] P. Wang, F. Casadei, S.H. Kang, K. Bertoldi, Locally resonant band gaps in periodic beam lattices by tuning connectivity, Physical Review B, 91 (2015).